%% file: 3drgcpaper.tex
\documentclass[onefignum,onetabnum]{siamart171218}

\usepackage{cite}

\usepackage{amsfonts}
\usepackage{bm}
\usepackage{graphicx}
\usepackage{epstopdf}
\usepackage{algorithmic}
\usepackage{subcaption}

\DeclareCaptionSubType*[Alph]{table}
\DeclareCaptionLabelFormat{mystyle}{Table~\bothIfFirst{#1}{ }#2}
\ifpdf
  \DeclareGraphicsExtensions{.eps,.pdf,.png,.jpg}
\else
  \DeclareGraphicsExtensions{.eps}
\fi

\newsiamremark{remark}{Remark}
\newsiamremark{hypothesis}{Hypothesis}
\crefname{hypothesis}{Hypothesis}{Hypotheses}
\newsiamthm{claim}{Claim}

\headers{Rigidity Percolation in Disordered 3D Rod Systems}{S. Heroy, D. Taylor, F. Shi, M. G. Forest, and P. J. Mucha}

\title{Rigidity Percolation in Disordered 3D Rod Systems
\thanks{This work was supported by the U.S. Army Research Office (\#W911NF-13-1-0013 and \#W911NF-16-1-0356), and by the National Science Foundation awards DMS-1664645 and CISE-1931516. Additional support was provided by a James S. McDonnell Foundation 21st Century Science Initiative - Complex Systems Scholar Award (\#220020315) and by the Eunice Kennedy Shriver National Institute of Child Health \& Human Development of the National Institutes of Health (R01HD075712).
The content is solely the responsibility of the authors and does not necessarily represent the official views of any of the agencies that supported this work.
}}

\author{Samuel Heroy\thanks{The Bartlett Centre for Advanced Spatial Analysis, University College London, London W1 4TJ, UK; Mathematical Institute, University of Oxford, Oxford OX2 6GG, UK
  (\email{sam.heroy@gmail.com})}
\and Dane Taylor\thanks{Department of Mathematics, University at Buffalo, State University of New York, Buffalo, NY 14260, USA}
\and Feng Shi\thanks{The Odum Institute for Research in Social Science, University of North Carolina, Chapel Hill, NC 27599, USA}
\and M. Gregory Forest\thanks{Carolina Center for Interdisciplinary Applied Mathematics, Department of Mathematics, University of North Carolina, Chapel Hill, NC 27599, USA}
\and Peter J. Mucha\footnotemark[5]}

\ifpdf
\hypersetup{
  pdftitle={Rigidity Percolation in Disordered 3D Rod Systems},
  pdfauthor={S. Heroy, D. Taylor, F. Shi, M.G. Forest, \& P.J. Mucha}
}
\fi

\begin{document}
\maketitle

\begin{abstract}
 \input{abstract.tex}
\end{abstract}



\input{introduction.tex}

\input{background.tex}

\input{methodology.tex}
\input{numerics.tex}

\input{analyses.tex}

\input{discussion.tex}

\appendix
\section{Rigidity motifs} 
\label{app:rigidity proofs}
\input{proofs}

\section{Algorithmic details for \emph{3D-RGC} and the effect of compression ordering} 
\input{algorithmicdetails.tex}
\label{app:compression}
\section{Contact percolation results}
\input{contact}
\section*{Acknowledgments}
We would like to thank Donald Jacobs for helpful comments and the chance to discuss this work with his research group at the University of North Carolina at Charlotte. We would also like to thank Theo J. Dingemans (UNC), Maruti Hegde (UNC), Ryan Fox (Exponent Inc), and Daphne Klotsa (UNC) for helpful collaborative conversations that helped motivate this work.

\bibliographystyle{siamplain}
\bibliography{references}
\end{document}

%% file: abstract.tex
In composite materials composed of soft polymer matrix and stiff, high-aspect-ratio particles, the composite undergoes a transition in mechanical strength when the inclusion phase surpasses a critical density. This phenomenon (rheological or mechanical percolation) is well-known to occur in many composites at a critical density that exceeds the conductivity percolation threshold. Conductivity percolation occurs as a consequence of contact percolation, which refers to the conducting particles’ formation of a connected component that spans the composite. Rheological percolation, however, has evaded a complete theoretical explanation and predictive description.  
A natural hypothesis is that rheological percolation arises due to rigidity percolation, whereby a rigid component of inclusions spans the composite. We model composites as random isotropic dispersions of soft-core rods, and study rigidity percolation in such systems. Building on previous results for two-dimensional systems, we develop an approximate algorithm that identifies spanning rigid components through iteratively identifying and compressing provably rigid motifs---equivalently, decomposing giant rigid components into rigid assemblies of successively smaller rigid components. We apply this algorithm to random rod systems to estimate a rigidity percolation threshold and explore its dependence on rod aspect ratio ($\alpha$). We show that this transition point, like the contact percolation transition point, scales inversely with the average ($\alpha$-dependent) rod excluded volume. However, the scaling of the rigidity percolation threshold, unlike the contact percolation scaling, is valid for relatively low $\alpha$. Moreover, the critical rod contact number is constant for $\alpha$ above some relatively low value; and lies below the prediction from Maxwell’s isostatic condition.

%% file: introduction.tex
\section{Introduction}
\label{sec:intro}
The application of the graph theoretic notion of rigidity towards studying material properties has a long and fruitful history, tracing back to Maxwell and finding place in present-day technologies. In the simplest conception (\emph{Maxwell counting} or \emph{Maxwell's isostatic condition}), a system of particles is predicted to be rigid when the number of inherent particles is met or exceeded by the number of constraints between them \cite{maxwell}. This criterion has two subtle shortcomings. First, it supposes that all constraints are independent. This has been corrected for in two-dimensional systems using \emph{Laman's condition}; more generally, \emph{rigidity matroid theory} offers a solution that recovers the global count of (linearly) independent constraints in systems of any dimension \cite{graver}. Second, this approach is directed towards the question of whether \emph{all} degrees of freedom in a system are bound---but in many applications (in which it is used), the more appropriate question is whether or not the system contains a \emph{giant} rigid component, i.e.\ a spatially extended, spanning subgraph that is rigid (\emph{rigidity percolation}). The first of these shortcomings underpredicts the rigidity percolation threshold, while the second overpredicts it---the relevance of Maxwell counting depends on a balance of these effects \cite{rigidity_fibers2d}. Nonetheless, this approach is used in many studies to characterize both simulations and laboratory experiments in various systems.

In recent years, rigidity theory has received considerable attention. In particular, relating the rigidity theory of glasses to chemical composition has aided in the development of Gorilla Glass \cite{gorilla_glass,gorilla_glass2}---a highly useful and lucrative component of mobile phones and other devices. Indeed, the application of rigidity theory to glasses has a rich history, tracing back to Phillips and Thorpe's simple constraint counting exercise \cite{chalcogenide,constraint_theory}. In a system of covalently bonded atoms (in three dimensions), interactions between atoms can be divided into (central force) two-body interactions and (bond-bending) three-body interactions which are shared between all pairs of second nearest neighbors. The sum of these constraints is given by $\frac{\langle c \rangle}{2}+(2\langle c \rangle-3)$, where $\langle c \rangle$ is the mean number of interactions per particle. Because the atoms each have three degrees of freedom, this leads to the isostatic condition $\langle c \rangle=2.4$, which is an estimate that has been shown to be quite accurate both in simulations \cite{glasses_rigidity_percolation,glasses_rigidity_diamond} and experiments, and has been used to predict the onset of critical mechanical behavior in chalcogenides \cite{chalcogenides_2,chalcogenides_3,chalcogenides_4}, oxide glasses \cite{oxide_1,oxide_2}, glassy metals \cite{glassy_metals}, and proteins \cite{proteins_1,proteins_2}. 

\begin{figure}[tp]
\begin{center}
\includegraphics[width=.8\linewidth]{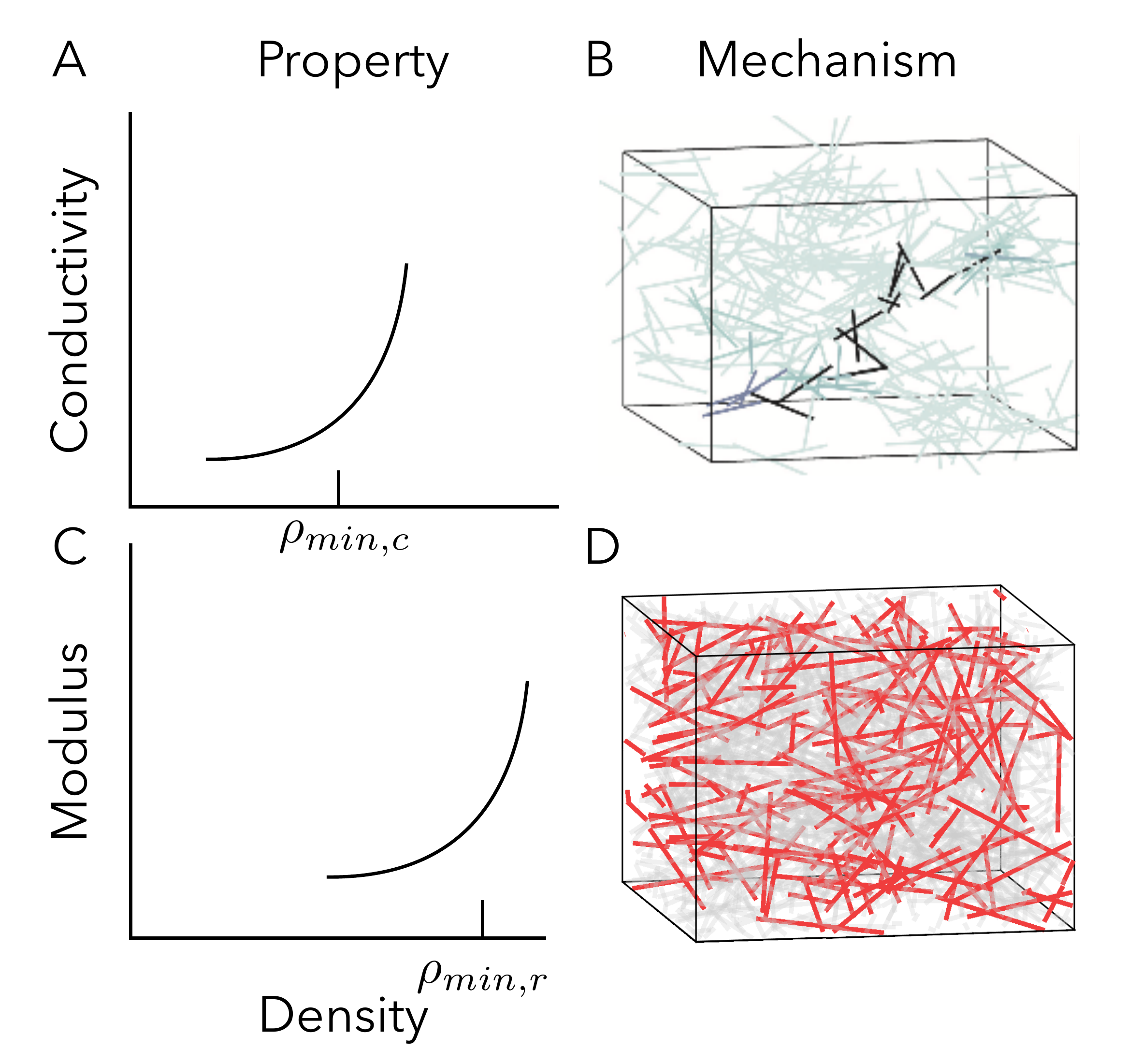}
\caption{\textbf{Conductivity percolation is driven by contact percolation, while rheological percolation is driven by rigidity percolation.} A--B: When the rod number density surpasses $\rho_{min,c}$, it becomes likely that a system contains a spanning component of rods, which --- if rod contacts are conducting --- leads to a nonlinear increase in the host system's conductivity. C--D: At a larger number density $\rho_{min,r}$, a giant rigid component emerges in the rod contact network, driving a transition in the mechanical stability of the material.}
\label{fig:schematic}
\end{center}
\end{figure}

Common amongst most systems in which the glass-forming condition is applied is the assumption that the inherent particles are `atoms' that are approximately spherical. Celzard \emph{et al.}\ apply this same approximation to derive the relationship between rigidity percolation and contact percolation\footnote{Contact/rigidity percolation have also been called  `scalar' and `vectorial' percolation in \cite{celzard} and other studies.} thresholds in systems of compressed expanded graphite \cite{celzard}. In particular, they load composites with increasing densities of inclusion particles - and measure conductivity/rheological percolation threshold as the (respective) inflection point in the observed relationship between the number density (number of particles per unit volume) of inclusion particles and the composite's electrical conductivity/mechanical stability (elastic modulus). Interestingly, though the inherent components are no longer simple atoms in this system, the glass-forming condition achieves success in predicting the ratio of the rheological percolation threshold to the conductivity percolation threshold. While conductivity (or electrical) percolation is frequently associated with contact percolation in the (conducting) rod phase \cite{shi,scaling}, Celzard \emph{et al.}\ \cite{celzard} constitutes the first study to our knowledge that uses the microscale phenomenon of rigidity percolation to predict the onset of rheological percolation in composite materials (see Fig.~\ref{fig:schematic}). 

However, while \cite{celzard} finds an experimental ratio of $1.7$--$1.8$ (relatively similar to the glass-forming prediction of $1.6$), other studies using various composite materials have found ratios as high as $3$ and as low as $1$, suggesting a need for further investigation \cite{penu}. Huisman and Lubensky apply a similar framework to a system of long filaments, which is constructed from one long filament that crosses itself 1000 times and is then subjected to a large number of Monte Carlo topology rewirings and segment cuttings. In this system---unlike in the glass-forming condition---only \emph{rod-sharing} second nearest neighbors share bond-bending constraints \cite{lubensky}. Using a similar argument to the glass-forming condition, they derive instead that $\langle z \rangle=\frac{10}{3}$ connections per junction (and $\rho_{min,r}/\rho_{min,c}=10/6$)\footnote{We use $\langle z \rangle$ here to refer to the average degree in a network wherein nodes represent contact points and edges represent the rod segments between contacts; and $\langle c \rangle$ to refer to the average degree in a network wherein rods/particles are represented by nodes and pairwise contacts correspond to edges.} is the appropriate isostatic condition for the system and demonstrate the efficacy of this prediction in simulations.  

Critical to Husiman and Lubensky's study \cite{lubensky} is the assumption that rods can rotate about but stay fixed to one another at their respective point of contact, i.e.\ the point of contact acts as a socket joint about which each rod can independently rotate. We can similarly derive Maxwell's isostatic condition in a rod-socket system\footnote{In two dimensions, positional (but not angular) constraint-inducing contacts between rods are equivalently hinge-like and socket-like. In that case, intersecting rods may rotate with one degree of freedom about the point of contact. We describe such contacts as socket-like in three dimensions, since the contacts allow intersecting rods two degrees of freedom about their respective contact point. Therefore, while the two-dimensional study \cite{rigidity_heroy} used the term ``rod-hinge system'' to describe the corresponding network, ``rod-socket system'' is more appropriate here in three dimensions.} in which rods are randomly dispersed in space and contact at intersections. Because each rod has $2D-1$ degrees of freedom in $D$ dimensions, and each contact constrains $D$ degrees of freedom, the system becomes rigid when $\langle c \rangle\ge \frac{4D-2}{D}$, where $\langle c \rangle$ is the average number of contacts per rod. In two-dimensional systems of random dispersed rods with socket/hinge-like contacts, the estimate $\langle c \rangle=3$ is rather inaccurate---the real value is about 4.25 \cite{rigidity_fibers2d,rigidity_heroy}---whereas it is quite accurate in the system of Husiman and Lubensky \cite{lubensky}. 

Several studies have studied rigidity percolation in three-dimensional rod-like systems (including \cite{lubensky,broedersz2011criticality,broedersz2014modeling}); however, none have characterized the problem in systems as the one we describe here (in which rods are randomly distributed and oriented in space). Understanding rigidity percolation in these systems is an important task because it is \emph{a priori} unclear whether any model created via lattice pertubations or long filament segmentations have sufficiently similar network topologies to systems of randomly positioned fibers. Moreover, while results in analogous 2D systems of randomly dispersed rods  can be used to qualitatively characterize experimental findings regarding rheological percolation, there are certain features of these networks that are unique to three dimensions. Although our approach also makes simplifications (e.g., that these rod dispersions are random and isotropic, and that contacts can be idealized as intersections), our study marks the first attempt to characterize rigidity percolation in 3D systems of randomly dispersed rods with socket-like contacts. 

In this study, we extend the \emph{rigid graph compression} (RGC) algorithm to three dimensional rod systems. We previously demonstrated that RGC can be used to approximate the rigidity percolation threshold with high accuracy in two-dimensional rod systems \cite{rigidity_heroy}. First (Sec.~\ref{sec:rgc}), we review the construction of this algorithm, which identifies components of mutually rigid rods through a set of hierarchical rules, and we develop the essential theory necessary to extend it to three dimensions (Sec.~\ref{sec:3dmeth}). Then, we use this algorithm to identify a rigidity percolation threshold in systems of randomly placed and oriented rods, conducting numerical experiments at various system sizes and aspect ratios and using a finite-size scaling analysis to estimate the transition and associated correlation length exponent (Sec.~\ref{sec:exp}). We use two approaches to assess the validity of this approximation and find that the approach robustly identifies a reasonably tight upper bound on the true threshold (Secs.~\ref{subsec:reshuffling}--\ref{subsec:accuracy}). Finally (Sec.~\ref{sec:interp}), we show that the rigidity percolation threshold (like the contact percolation threshold in the slender body limit) varies with the reciprocal of the excluded volume and generally lies below Maxwell's isostatic condition.

%% file: background.tex
\section{Background}
\label{sec:rgc}
In this section, we define the mathematical problem of interest to this paper---identifying rigidity percolation in systems of overlapping capped cylinders or sphereocylinders (`rods') that stay fixed at but can pivot about their contact points (`socket joints')---and outline previous work related to this topic in two and three dimensional systems. 
\subsection{Rigidity percolation and rigidity detection algorithms}
Suppose that $N$ rods of aspect ratio $\gamma=\ell/2r,$ where $\ell$ gives the lengths of their core cylinders and $r$ the radii, are positioned in a finite $D=3$ dimensional box of depth $L\gg\ell$ (i.e., volume $V=L^3$) such that both their centers and orientations are drawn from a uniformly random probability distribution (we defer the study of sequentially packed rods to future research). We allow the box to have periodic boundary conditions on all sides, meaning a rod intersecting the box in one dimension(s) will appear on the opposite end in the same dimension(s). We allow the rods to intersect one another, and associate each pairwise intersection with a single contact point that constrains the relative motions of the rod pair. That is, intersecting rods may individually rotate about but remain connected to one another at their shared contact point in accord with experimental findings that CNTs, for example, form interconnected networks with bonds that freely rotate and resist stretching \cite{hough2004viscoelasticity}. We call the resultant system a \emph{rod-socket system}. Our general problem is to identify for all $L$, the critical number density $\rho_{min,r}=N/L^D$ above which these rods will with high probability form a subgraph that is \emph{spanning} (intersecting both sides of the box domain along one dimension) and \emph{rigid} (having no internal degrees of freedom). We additionally wish to determine the scaling of the size of the largest rigid component for $\rho>\rho_{min,r}$, as well as the dependence of both the transition threshold and the scaling exponent on the rod aspect ratio, $\gamma$. 

In our study, we borrow from insights that we learned from studying the same problem in analogous 2-dimensional systems. In particular, we know of at least four methods that have been used to understand rigidity percolation in this setting. We have already described Maxwell's isostatic condition (which is a global mean field estimate), as well as its shortcomings. A second method relies on a spring-based relaxation algorithm that involves joining intersection points along shared rods via springs. When the positions of these intersection points are perturbed but then allowed to relax, the evolution of the system will recover initial distances (resting lengths) between points along shared rods, and in theory it will recover the initial distances between intersection points on rods in a mutually rigid cluster, while randomizing distances between intersection points in separate rigid clusters (alternatively, an SVD decomposition of the resulting matrix can be employed to the same effect \cite{pellegrino1993structural}). While the relaxation algorithm is generalizable to a variety of systems and applications \cite{lubensky,wilhelm2003elasticity,broedersz2011criticality,broedersz2014modeling}, its implementation in rod-socket systems is numerically unstable, though it has been shown to be tractable in the two-dimensional setting \cite{wilhelm2003elasticity}.\footnote{Our own efforts to apply this method in 3 dimensions have been unsuccessful so far.} A third method used for rigid cluster detection originates from a graph theoretic theorem, Laman's condition \cite{laman}, which gives a local combinatorial condition for rigidity in two-dimensional planar graphs that can be utilized algorithmically (the `pebble game' \cite{pebble}) in large systems and can be adapted to 2D rod-hinge/socket systems \cite{rigidity_fibers2d}. Unfortunately, this condition does not apply exactly to three dimensions, though it can be accurate in certain 3D systems, including bond-bending networks \cite{jacobs1998generic} (which are frequently used to model proteins \cite{jacobs2001protein}) and perturbed lattices \cite{threedpebble}; but the problem of combinatorial characterization of rigidity of graphs (and in particular rod-socket systems) in 3D remains open. See also \cite{sitharam2004tractable,pardalos2018open}.

While these latter two methods (spring relaxation and the pebble game) may at some point be utilized to characterize rigidity percolation in our system of interest, our contribution here lies in extending a fourth method, rigid graph compression (RGC), from 2D \cite{rigidity_heroy} to 3D. Like the pebble game, RGC utilizes a graph of rod connections to identify rigid components (rather than explicitly using the location of the rods/intersections in space, as in spring relaxation) and thereby determines if a system has a spanning rigid component. While the pebble game approximation identifies certain floppy components as rigid in 3D \cite{chubynsky2007algorithms} (it is exact in 2D), RGC exclusively identifies rigid components in both 2D and 3D. However, RGC may fail to identify some rigid components, and it therefore gives a sufficient but not necessary condition for identifying a subgraph as rigid or not. In contrast, the 3D pebble game approximation gives a necessary but not sufficient condition. In the remainder of this section, we describe the formulation of RGC in detail and highlight the results of its implementation in 2D. We then extend RGC to 3D in the next section.

\subsection{Rigidity theory for rod-socket systems}

Before discussing the application of rigidity matroids to our system, we briefly frame our model system in light of the wider study of rigidity theory. Most work in this general topic relates to rigidity of \emph{graphs} that can be described by pairwise constraints that fix distances between nodes (also called bar-joint networks with central force constraints and bond-bending constraints if second nearest neighbors have angular constraints) \cite{thorpe1999rigidity}. While we rely on these developments (which we discuss in the subsection to follow) in setting up our own methodology, our system differs in that individual particles are rods --- while there are pairwise constraints between points on the rods (e.g.\ endpoints or contact points), certain peculiarities of the system necessitate specialized treatment. 

Our work here is somewhat parallel to the study of 3D body bar networks (we refer interested readers to \cite{thorpe1999rigidity}), but our system has certain properties that these networks do not accommodate. Body bar networks are described by ``bar" constraints between rigid bodies that have the same dimension as the system but arbitrary shape (these are often used to represent biomolecules \cite{whiteley1999rigidity,hespenheide2004structural}). Whereas most analyses of body-bar networks assume contacts are noncollinear, rods often have collinear contact points, which in certain cases form redundant constraints. Moreover, these bodies are generally treated having the same dimension as the system, and lack any special symmetry (unlike axisymmetric rods). Similarly, one may be tempted to relate our rod system to body-hinge systems, which are described by hinges that connect along $D-2$ dimensions, but the hinges in 3D-space are line-hinges as opposed to point-sockets in our system (and again, ``bodies'' in our system are rods that necessitate special treatment) \cite{jackson2008pin,whiteley1996some}.

\subsection{Rigidity matroid theory for spatially embedded graphs}

RGC depends explicitly on being able to prove that certain contact patterns between components, which are themselves rigid, can be used to identify a larger rigid component that contains these individual rigid components. In particular, we use a dynamical systems framework in rigidity matroid theory to derive these rigid contact patterns or rigid motifs. Rigidity matroid theory uses a graph's embedding in Euclidean space, or `framework', $\alpha(G)$,\footnote{We note that in other papers, $\rho$ is frequently used to denote a framework. Instead, we reserve $\rho$ to denote number density.} to characterize its rigidity through the language of linear algebra \cite{singer,graver, hendrix}. Consider the set of node positions of some graph $G(V,E)$ to be a dynamical system such that $\bm p_i(t)$ is the $D$-dimensional position of node $i$ at time $t.$ The condition that each edge $e_{ij}\in E$ maintains a fixed distance $d_{ij}$ between nodes $i$ and $j$ requires $||\bm p_i(t)-\bm p_j(t)||_2^2=d_{ij}^2\;\;\forall e_{ij}\in E$.
Since this quadratic system is not computationally convenient, it is common to linearize by differentiating with respect to time, obtaining
\begin{equation}
(\bm p_i(t)-\bm p_j(t))\cdot(\bm u_i(t)-\bm u_j(t))=0
\qquad
\forall e_{ij}\in E\,,
\label{rmt}
\end{equation} 
where $\bm u_i(t)={d\bm p_i(t)}/{dt}$ is the instantaneous velocity of node $i$. The totality of these constraints informs an $|E|\times D|V|$ matrix, $\bm X$---the rigidity matrix of $\alpha(G)$---satisfying $\bm X \bm u= \bm 0$, where $\bm u$ is the $D|V|$-vector of velocities. A vector $\bm u$ satisfying $\bm X \bm u= \bm 0$ is an \emph{infinitesimal motion} of $\alpha(G)$, and the right nullspace of $\bm X$ includes the full set of such motions. 
If $G$ is embedded in Euclidean $\mathbb{R}^D$ and the right nullspace of $\bm X$ spans only the $D(D+1)/2$ rigid-body motions of translation and rotation, the framework $\alpha(G)$ is said to be \emph{infinitesimally rigid}. Otherwise, $\alpha(G)$ is \emph{infinitesimally flexible}.

Importantly, it has been shown generically that if a framework $\alpha(G)$ is infinitesimally rigid, then almost all other realizations of $G$ are infinitesimally rigid, the exceptions of which form a set of measure zero \cite{asimow1978rigidity}. Therefore, one can (generically) infer rigidity from the topology of a graph itself, rather than from any particular embedding in space (Sec. 3 of \cite{asimow1978rigidity}).
We note that this argument breaks down when at least one nontrivial minor of $\bm X$ has a zero determinant---however, these cases occur with probability zero in random systems. Practically, determining the rigidity of $\alpha(G)$ thus reduces to computing the rank of $\bm X$, and using the rank nullity theorem to then determine the dimension of the matrix's nullspace, which corresponds injectively to the count of the underlying graph's degrees of freedom.

\subsection{Rigidity matroid theory for interacting rigid components}
\label{subsec:interacting}
While it may seem appealing to use rigidity matroid theory to capture rigidity percolation, computational rank estimation is subject to numerical difficulties and, more importantly, the rank calculation can only be used to give a system count for the number of degrees of freedom, rather than finding a spanning rigid component. In our prior work (Sec. 3.1. of \cite{rigidity_heroy}) we instead proposed to use rigidity matroid theory to study the motions of small numbers of interacting rigid components, or rigid motifs. These rigid motifs are essential ``building blocks" that hierarchically compose giant or spanning rigid motifs. We review our study of rigid motifs here and in the next subsection. 

The motions of any $D'$-dimensional rigid component in $D$ dimensions can be fully determined from $D'+1$ points contained in the component, the translations and rotations of which together generate the Special Euclidean group \emph{SE(D)} \cite{geometries}. (In principle, fewer coordinates are needed if employing angular constraints, but for simplicity we work with $D'+1$ points.) Hence, for some rigid body $R\subset\mathbb{R}^{D'}$, which we define as the union of volumes enclosed within some integer number $r>0$ of 1-dimensional rods, we affix a \emph{coordinate labeling} of $R$, composed of either $D'+1=2$ (singleton rod)\footnote{Note that regardless of the dimension of the rod network, if $R$ includes only a single rod, then it has a spatial dimension $1$, and so only two points are needed to specify its rigid motions.}, $3$ (planar), or $4$ (nonaxisymmetric, nonplanar) points $\{\bm p_i\}$ which fully capture the rigid motions of $R$. Importantly, no more than two points in a coordinate labeling may be collinear, or else the coordinate labeling will only capture a subset of the available rigid motions of the corresponding body. Due to the rigidity of $R$, the pairwise distances between the points are fixed, providing ${D'+1}\choose{2}$ constraint rows in the corresponding rigidity matrix $\bm X$, each having the form:
\begin{equation} 
\Delta \bm p_{i,j} \cdot \bm u_{i} -\Delta \bm p_{i,j} \cdot \bm u_{j}=0, 
\label{eq:rigidity_row}
\end{equation} 
where $\bm u_i$ and $\bm u_j$ are the instantaneous velocities corresponding, respectively, to the points $\bm p_i$ and $\bm p_j$ (each of dimension $D$) that are affixed in $R$ and $\Delta \bm p_{i,j}=\bm p_i-\bm p_j$. 

When two or more rigid components $R_1,\,R_2,\,\dots$ are in contact, we denote the composite system by $R_1* R_2*\dots$ and the corresponding composite rigidity matrix by $\bm X_1 * \bm X_2*\dots$. For such systems, we construct a \emph{minimal coordinate labeling}, defined as the union of coordinate labelings for each involved $R_i$ such that coordinate labelings include interaction points between rigid components wherever possible. For a given set of coordinate labelings of all included rigid components, we construct a \emph{constraint graph} encoding the topology of physical constraints between the rigid components (Fig.~\ref{fig:visuals} here or Fig. 3.1 in \cite{rigidity_heroy}). This graph is constructed by creating for each rigid component $R_i$ with $S_i$ 
coordinate labels a $(S_i)$-clique---that is, an all-to-all connected subgraph. The constraint graph is defined as the union of these cliques.
\subsection{Rigid motifs and RGC}
\label{subsec:2drgc}
In \cite{rigidity_heroy} and in the present study, we developed a methodology to use rigidity matroid theory - supplemented with the used of coordinate labelings - to develop rules for aggregating 2D rods (line segments) and non-axisymmetric composite rigid bodies into larger composite rigid bodies. These rules are expressed as \emph{primitive rigid motifs}, which represent topological `building blocks' of rigidity in rod-socket systems. We use the term `primitive' because these motifs may not be decomposed into simpler motifs, yet many larger, more complicated patterns of interaction can be constructed from these motifs, analogous to the formation of Laman graphs from Henneberg constructions (in bar-joint systems) \cite{henneberg1911graphische,tay1985generating}. In \cite{rigidity_heroy} and the present work, rigid components will be rods or sets of connected rods, but the formalism does not necessarily require the individual particles to be rod-shaped, and the methodology may with only slight modification be extended to ellipsoids, curvilinear filaments, and other axisymmetric shapes, so long as the interactions between the particles are socket-like. 

In \cite{rigidity_heroy} and here, we use these primitive rigid motifs to identify large-scale rigid components agglomerated from rigid components identified at smaller scales, starting from the microscopic scale of primitive rigid motifs acting on individual rods  (Fig. 3.2 in \cite{rigidity_heroy} and Fig.~\ref{fig:3d motifs} here). Our approach relies on constructing rod contact networks in which each node represents a rigid component and edges indicate which components intersect with one another (similar to body bar networks). This network construction contrasts the constraint graphs described earlier (in which nodes represented coordinate labelings and edges represent rigidity constraints).

In \cite{rigidity_heroy}, we leverage these rigid motifs into a graph-compression algorithm (RGC) that decomposes large rod-socket systems into their rigid components. The RGC algorithm involves initialization (Step 1), followed by iterative identification and compression of rigid motifs (Steps 2 \& 3): 
\begin{enumerate}
    \item given a set of interacting particles (i.e.\ a rod-socket system), construct a contact network of rods represented as nodes and contacts between rods represented as edges;
    \item identify rigid motifs in the contact network;
    \item compress each rigid motif instance into a single node, yielding a reduced set of rigid components and an updated contact network;
    \item return to Step 2.
\end{enumerate}
This algorithm terminates when no more rigid motifs can be identified. In \cite{rigidity_heroy}, we show that RGC is a useful framework for identifying rigidity percolation in 2D rod-hinge/socket systems, and in particular that an implementation using only 3 rigid motifs can identify the rigidity percolation threshold to within $0.6\%$ relative error. Moreover, we show that the 2D algorithm is robust to the order in which motifs are identified/compressed---i.e.\ the order has no effect on the final rigid components identified.

%% file: methodology.tex
\section{RGC in three dimensions}
\label{sec:3dmeth}
Our main contribution herein is to extend RGC to 3D systems. As the application of interest (rheological percolation) occurs generally in 3D systems, it is of paramount interest to understand rigidity percolation in this more complex domain. In Sec. 3.1 below, we first discuss methodological challenges facing such an extension. We then present 3D rigidty motifs in  Sec. 3.2 and our 3D RGC algorithm in Sec. 3.3. 

\subsection{Methodological challenges in 3D}
\label{subsec:challenges}
The core approach of RGC in 2D and 3D is essentially the same---in a graph of connections between interacting rods, we identify specific motifs and compress these based upon physically meaningful rules (rigid motifs) until no more can be identified. However, we can identify at least four challenges that complicate the 3D approach, which we enumerate here:
\begin{enumerate}
\item In 2D, rods and planar rigid bodies have the same number (3) of degrees of freedom. However, rods (and other axisymmetric rigid bodies) have 5 distinguishable degrees of freedom in 3D while non-axisymmetric rigid bodies have 6. Therefore, the rigid motifs that we identify must discriminate between the two, and the algorithm must keep track of which nodes represent rods and which do not as the graph compresses. (We note that composite rigid bodies constructed from multiple rods are always non-axisymmetric.) 
\item In 2D, it was only necessary to consider interactions in which pairs of rigid bodies share $\le 2$ contacts. Owing to the increased number of degrees of freedom for rigid bodies in 3D, this will not always be the case in the rigid motifs of the next subsection. As noted in Sec.~\ref{subsec:interacting}, a coordinate labeling must be chosen such that no subset of $\ge 2$ contained points are collinear because such collinearities will lead to linear dependencies in the associated rigidity matrix. However, if $\ge 2$ contacts occur along a single  rod (i.e., if they lie along the axis of the rod), then the same number are collinear in our treatment. Therefore, in treating compositions between rigid bodies sharing $\ge 2$ contacts in 3D, it is critical to check whether or not these contacts are rod-sharing. If $\ge 2$ contacts are not rod-sharing, we assume under generic conditions that they are not collinear. At the same time, we assume all rod-sharing contacts to be collinear. Strictly speaking, rod-sharing contacts may not necessarily be collinear if they do not lie along the same axis of a finite aspect ratio rod. However, for high-aspect ratio rods, they will in most cases be nearly collinear. in order to establish an upper bound for the rigidity percolation threshold, we assume any rod-sharing contacts to be collinear. 
\item In [23], we showed for RGC algorithms with three primitive motifs  that different orderings of these specific motifs in RGC would yield the same identification of rigid components upon compression. Moreover, rigid clusters in 2D ``come in one piece'' \cite{threedpebble}, meaning that no rod could be a member of more than one rigid cluster. Neither of these conveniences are true in 3D, as we show by counterexample in Appendix~\ref{app:compression}. 
\item For 2D rod-hinge systems, we were able to compare the performance of RGC to exact rigidity analyses based on the pebble game cite \cite{rigidity_heroy,rigidity_fibers2d}. However, for our study of disordered 3D systems, there does not currently exist such a ``ground truth'' with which to compare.
\end{enumerate}

\subsection{Rigid motifs in 3D}
\label{subsec:3dmotifs}
Here, we present seven rigid motifs that we will utilize to design \emph{3D-RGC} algorithms for identifying rigid components in 3D. To simplify our discussion, we adopt the naming schema ``Motif xDyz" with x indicating the spatial dimension ($3$), y indicating the number of aggregating rigid components in the motif, and z indicating an alphabetical index if there are different motifs associated with the same x and y.  We prove in this subsection that Motifs 3D2A and 3D2B are rigid and outline the similarly constructed proofs for the various other motifs in Appendix~\ref{app:rigidity proofs}. In Fig.~\ref{fig:visuals}, we offer a visual guide to these proofs by depicting an exemplifying physical rod network, coordinate labeling, and constraint graph for each respective motif.

As in two dimensions, we assume generic conditions in which no pair of rods is precisely parallel. In three dimensions, contacts are not points but are rather three-dimensional volumes. Therefore, when designing these rules, we specify how many \emph{instances} in which the relevant rigid components intersect. By an instance, we precisely mean the unique intersection of two rods in the physical rod network. While two non-axisymmetric rigid components may intersect in any number of instances, and the same is true for one non-axisymmetric rigid component and one rod, pairs of rods only intersect in one instance. Our strategy is then to choose a point within each contact intersection. If three or more contacts lie within the same rod (we call these rod-sharing from here on), we treat the corresponding points as precisely collinear except in one special case. In the special case in which three or more rigid components intersect in the same volume, we choose corresponding points such that these are not collinear. Finally, we use set notation $\cap,\cup,/$ to indicate intersections, unions, and set differences between the volumes in respective rigid components. For example, $R_1\cup R_2$ gives the set of all volumes enclosed in the rods composing $R_1$ and $R_2,$ while $R_1\cap R_2$ gives the volumes corresponding to any intersections between the rods composing $R_1$ and the rods composing $R_2$.
\begin{figure}[htbp]
\begin{center}
\includegraphics[width=.8\linewidth]{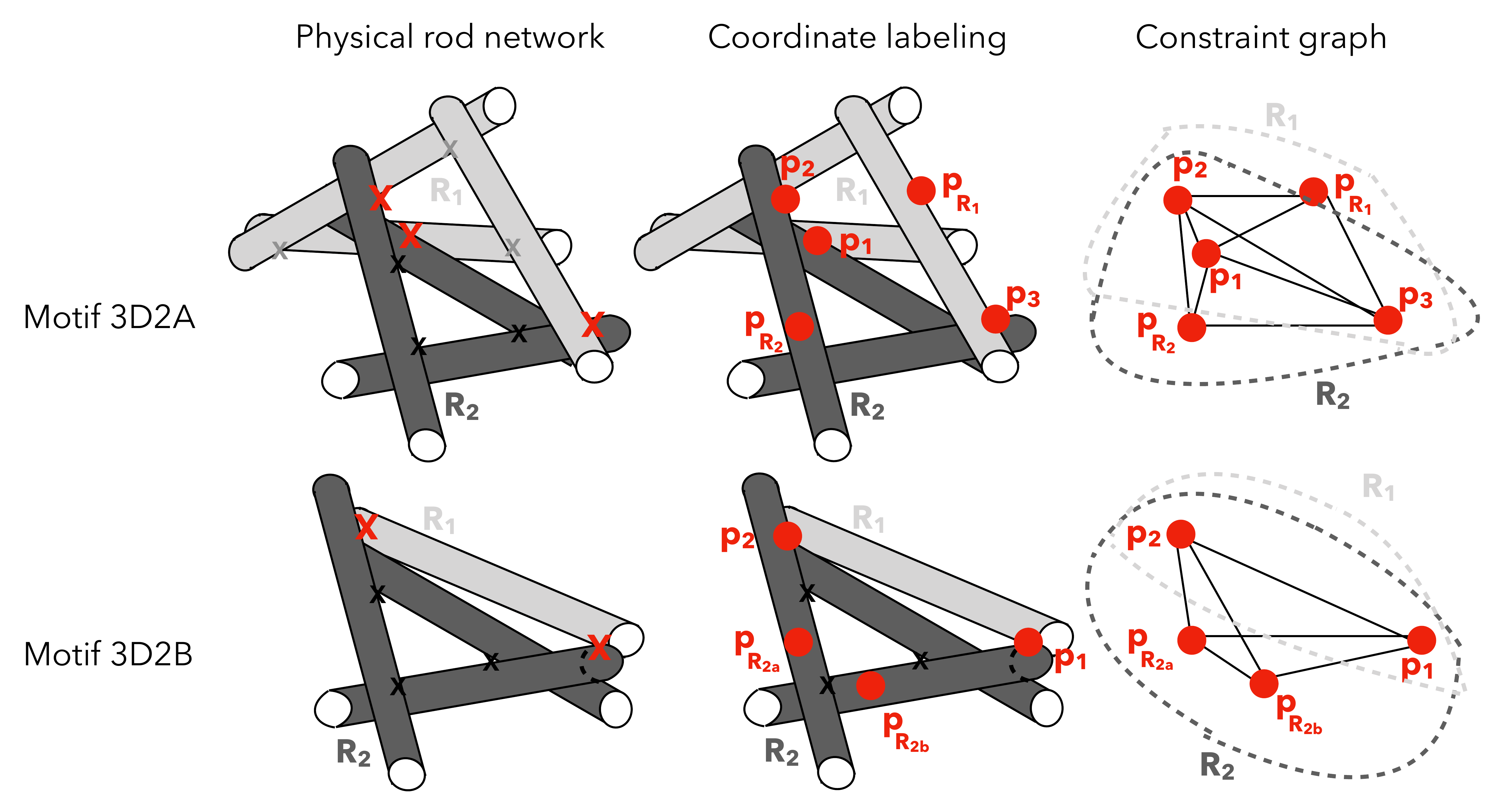}
\caption{Two rigid motifs in three dimensions. \emph{Left column:} Rigid components (distinguished here by shade), which may be individual rods or sets of connected rods, intersect with two specific topologies as described in Sec.~\ref{subsec:3dmotifs} to form larger-scale rigid components: (top) two non-axisymmetric rigid components $R_1$ and $R_2$ interact at three noncollinear points; (bottom) two rigid components, one axisymmetric (a rod $R_1$) and the other ($R_2$) non-axisymmetric, interact at two points. The contacts within individual rigid components are indicated by small gray and black x's, whereas the contact between them are indicated by large red x's. \emph{Middle column:} Coordinate labelings are affixed to each rigid component: four noncollinear points are required to describe the motions of a 3D rigid component consisting of multiple rods, whereas individual rods are 1D and require only two points. For each motif, we identify a set of minimal coordinate labelings that include intersection points whenever possible (see text for clarification). \emph{Right column:} The coordinate labelings give rise to constraint graphs in which edges (black lines) indicate fixed distances between the points they connect. The dashed ellipses group the points according to which rigid component they belong (i.e., $R_1,R_2$). By Theorem~\ref{motif 3D2A2B}, these constraint graphs, and the motifs that generate them, are rigid in three dimensions.}
\label{fig:visuals}
\end{center}
\end{figure}

\begin{theorem}
\label{motif 3D2A2B}
(Rigidity of Motif 3D2A and Motif 3D2B) The composition of two non-axisymmetric rigid components $R_1$ and $R_2$ that intersect at three or more instances, which are not all rod-sharing, is rigid in three dimensions (3D2A). If one of these components is axisymmetric (say, $R_1$), then only two of these intersections ($\bm p_1, \bm p_2$) are necessary (3D2B).
\end{theorem}
\begin{proof}
We first show that motif 3D2A is rigid. Let $\bm p_1,\bm p_2,\bm p_3$ be points in each of the intersections between the two bodies ($R_1\cap R_2$).  Let us first assume that these non-axisymmetric rigid bodies are nonplanar. In this case $R_1$ and $R_2$ have four points apiece in their coordinate labelings. We choose three of these points in each labeling to lie respectively in the pairwise intersections, giving the respective coordinate labelings $\{ \bm p_1, \bm p_2, \bm p_3, \bm p_{R_1}\}$ and $\{ \bm p_1, \bm p_2, \bm p_3, \bm p_{R_2}\}$, where $\bm p_{R_1}$ is a point in $R_1\backslash R_2$ and vice versa. Such selections give the rigidity matrices for $R_1$ and $R_2$:
\begin{align}
\bm X_1 =\left[
\begin{array}{ccccc}
\Delta \bm p_{1,2}&-\Delta \bm p_{1,2} &\bm 0&\bm 0&\bm 0\\
\bm 0 &\Delta \bm p_{2,3}&-\Delta \bm p_{2,3}&\bm 0&\bm 0\\
\Delta \bm p_{1,3}&\bm 0 &-\Delta \bm p_{1,3} &\bm 0&\bm 0\\
\Delta \bm p_{1,R_1} & \bm 0& \bm 0 &-\Delta \bm p_{1,R_1} &\bm 0\\
\bm 0 & \Delta \bm p_{2,R_1} &\bm 0 &-\Delta \bm p_{2,R_1}&\bm 0\\
\bm 0 &\bm 0&\Delta \bm p_{3,R_1}&-\Delta \bm p_{3,R_1} &\bm 0\\
\end{array}\right],\\
\bm X_2 =\left[
\begin{array}{ccccc}
\Delta \bm p_{1,2}&-\Delta \bm p_{1,2} &\bm 0&\bm 0&\bm 0\\
\bm 0 &\Delta \bm p_{2,3}&-\Delta \bm p_{2,3}&\bm 0&\bm 0\\
\Delta \bm p_{1,3}&\bm 0 &-\Delta \bm p_{1,3} &\bm 0&\bm 0\\
\Delta \bm p_{1,R_2}&\bm 0&\bm 0&\bm 0&-\Delta \bm p_{1,R_2}\\
\bm 0&\Delta \bm p_{2,R_2}&\bm 0&\bm 0&-\Delta \bm p_{2,R_2}\\
\bm 0&\bm 0&\Delta \bm p_{3,R_2}&\bm 0 &-\Delta \bm p_{3,R_2}
\label{eq:x1_and_x2}
\end{array}\right].
\end{align}
Note that $\bm X_1$ and $\bm X_2$ are each of size $6\times15$, and each column represents infinitesimal motion of a point in $\mathbb{R}^3$. Note that there are 6 rows in each rigidity matric, since there are ${4 \choose 2}=6$ possible pairwise constraints between four points. For the composite system, there are 9 rows since there are ${5 \choose 2} =10$ possible pairwise constraints between 5 points, however there is no constraint between points $\bm p_{R_1}$ and $\bm p_{R_2}$. The composite $9\times 15$ rigidity matrix $\bm X_1\ast \bm X_2$ is given by the intersection of these rows (taking row permutations where convenient):
\begin{align}
\bm X_1 \ast \bm X_2=\left[
\begin{array}{ccccc}
\Delta \bm p_{1,2}&-\Delta \bm p_{1,2} &\bm 0&\bm 0&\bm 0\\
\bm 0 &\Delta \bm p_{2,3}&-\Delta \bm p_{2,3}&\bm 0&\bm 0\\
\Delta \bm p_{1,3}&\bm 0 &-\Delta \bm p_{1,3} &\bm 0&\bm 0\\
\Delta \bm p_{1,R_1} & \bm 0& \bm 0 &-\Delta \bm p_{1,R_1} &\bm 0\\
\bm 0 & \Delta \bm p_{2,R_1} &\bm 0 &-\Delta \bm p_{2,R_1}&\bm 0\\
\bm 0 &\bm 0&\Delta \bm p_{3,R_1}&-\Delta \bm p_{3,R_1} &\bm 0\\
\Delta \bm p_{1,R_2}&\bm 0&\bm 0&\bm 0&-\Delta \bm p_{1,R_2}\\
\bm 0&\Delta \bm p_{2,R_2}&\bm 0&\bm 0&-\Delta \bm p_{2,R_2}\\
\bm 0&\bm 0&\Delta \bm p_{3,R_2}&\bm 0 &-\Delta \bm p_{3,R_2}
\end{array}\right].
\label{3daxiom1b}
\end{align}
Consider that it is a block triangular matrix with diagonal blocks: 
\begin{equation}
\left \{
\left[\begin{array}{cc} \Delta \bm p_{1,2} &-\Delta \bm p_{1,2} \end{array}\right],\left[\begin{array}{c} -\Delta \bm p_{2,3}\\ -\Delta \bm p_{1,3} \end{array}\right],\left[\begin{array}{c} -\Delta \bm p_{1,R_1}\\ -\Delta \bm p_{2,R_1}\\ -\Delta \bm p_{3,R_1} \end{array}\right],\left[\begin{array}{c} -\Delta \bm p_{1,R_2}\\ -\Delta \bm p_{2,R_2}\\ -\Delta \bm p_{3,R_2} \end{array}\right]
\right \},
\label{eq:3d2_reduced}
\end{equation}
Because the interaction points of $R_1$ and $R_2$ are chosen to be noncollinear, and because we have excluded the case that the intersection points are all rod-sharing (collinear), these blocks have ranks $1$, $2$, $3,$ and $3,$ respectively. Therefore, the block triangular matrix and the original composite rigidity matrix are both full rank, i.e.\ $\text{rank}(\bm X_1\ast \bm X_2)\ge 9$ and $\text{dim}(\text{null}(\bm X_1\ast \bm X_2))\le6$. Since a rigid body in three dimensions (lacking any symmetries) has six degrees of freedom, we conclude that $R_1\cup R_2$ is rigid, where we use $\cap/\cup$ notation to indicate the union or intersection of all rods in the respective bodies.

Now, suppose one of the two rigid bodies (say, $R_1$) is planar (as in Fig.~\ref{fig:visuals})---in this case, only three points are necessary to specify its coordinate labeling. Then, rows $4$--$6$ in Eq.~\ref{3daxiom1b} may be dropped, as can columns $10$--$12$ (recalling each entry in the displayed matrix represents a $1\times 3$ block), leading to block diagonalization that is equivalent to Eq.~\ref{eq:3d2_reduced}, except that the third block is dropped. Finally, these linearly independent diagonal blocks have ranks that sum to 6, while the column space has dimension 12, so the matrix again has rank 6 and is rigid. The case in which both $R_1$ and $R_2$ are planar follows similarly.

Next, we turn to the motif (Motif 3D2B) in which one of the rigid bodies is a rod (say $R_1$) that intersects the non-axisymmetric rigid body $R_2$ at two or more points $\bm p_1, \bm p_2,\dots$. In this case, we simply choose $\bm p_1$ and $\bm p_2$ as the coordinate labeling for $R_1$---and, if the body $R_2$ is nonplanar, we choose its coordinate labeling $\{ \bm p_1, \bm p_2, \bm p_{R_{2a}}, \bm p_{R_{2b}}\}$ where the latter two points are chosen such that no three member subset of the labeling is collinear. If instead $R_2$ is planar, only one of these latter points is necessary. In either case, the composite rigidity matrix is exactly the same as the rigidity matrix $\bm X_2$, which is rigid by hypothesis.
\end{proof}
The other rigidity motifs that we study are visualized via their contact network representations in Fig.~\ref{fig:3d motifs}, and are proven in Appendix~\ref{app:rigidity proofs}.

\subsection{Implementation of RGC in 3D}
\label{sec:method}
\begin{figure}[htbp]
\begin{center}
\includegraphics[width=.8\linewidth]{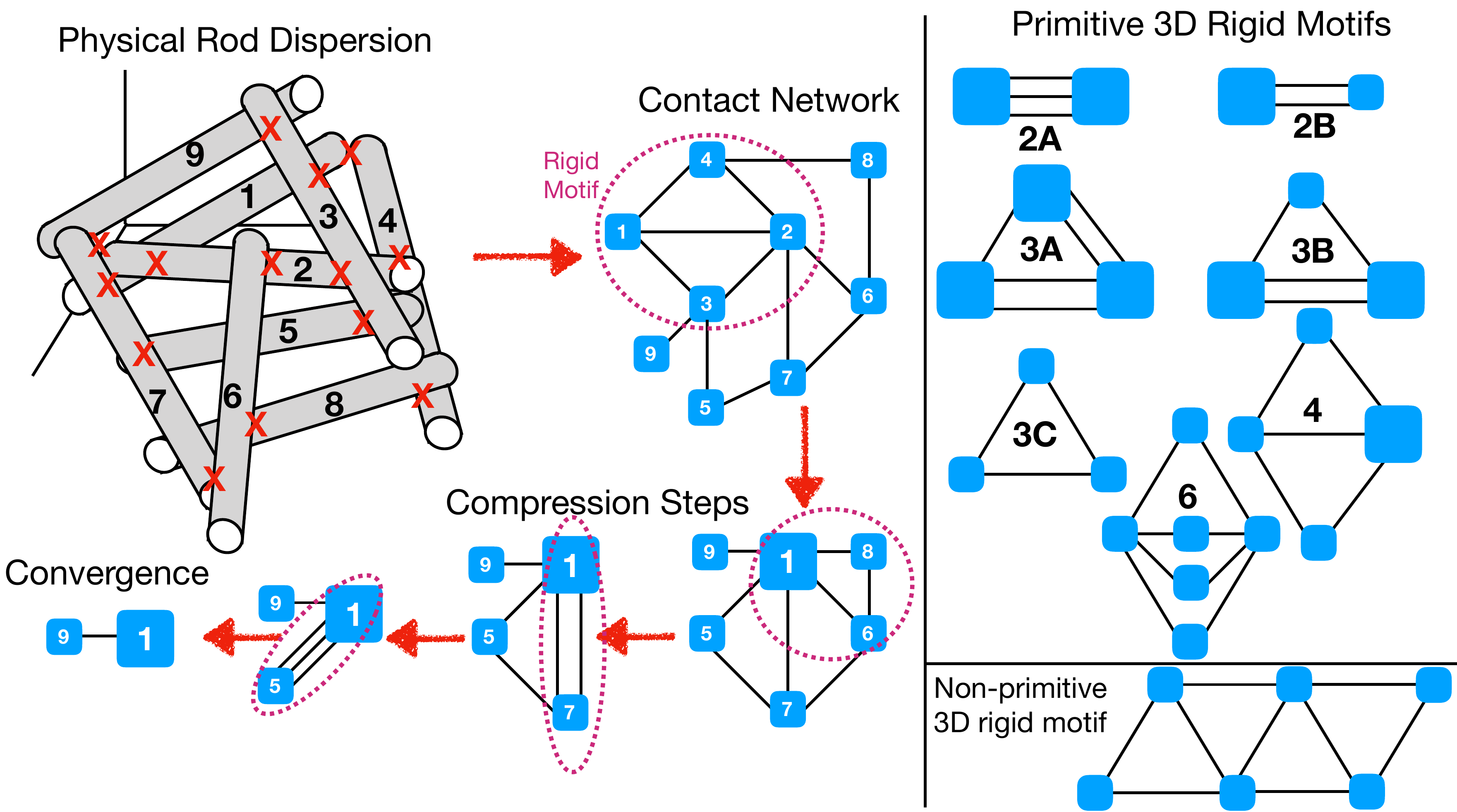}
\caption{\emph{Left}: In each Monte Carlo simulation, $N_R$ sphereocylinder rods are randomly placed in a periodic box ($N_R=9$ in the visualization here). From these rods, a contact network is extracted in which rods are represented as nodes (small blue squares) which share edges if the underlying pair have a physical intersection. Then, the RGC algorithm is used to find and iteratively compress the network's rigid components---sets of nodes that are provably mutually rigid---into single nodes (larger blue squares). This algorithm proceeds until convergence, which is attained when the algorithm has achieved a minimal decomposition of the network into components of mutually rigid rods (i.e., until no rigid motifs are identifiable). Here, rods $1$--$8$ form one rigid component, and rod $9$ is a second rigid component. (Note that rod 9 can rotate about its contact point in this component). \emph{Right}: We prove in Sec.~\ref{subsec:3dmotifs} and Appendix~\ref{app:rigidity proofs} that these rigid motifs are indeed contact patterns that fully constrain their inherent rods or non-axisymmetric rigid bodies.}
\label{fig:3d motifs}
\end{center}
\end{figure}

Aside from the difficulties described in Sec.~\ref{subsec:challenges}, implementation of RGC in 3D (Fig.~\ref{fig:3d motifs}) is a straightforward extension of the 2D implementation that we described in Sec.~\ref{subsec:2drgc} and in \cite{rigidity_heroy}. However, we make a few practical notes here.  Rather than identifying instances of Motif 3D3B directly, we make use of an available fast algorithm for identifying $k$-clique communities, which are sets of k-cliques (complete subgraphs on $k$ nodes) that are joined pairwise at $k - 1$ points \cite{cpm,networkx}. In particular, any $3$-clique community of rods (see Fig.~\ref{fig:3d motifs}) is necessarily a composite rigid motif in 3D, as it can be generated via repeated application of Motif 3D2B to Motif 3D3C. Furthermore, every instance of Motif 3D3C (triangular arrangements of rods) is a member of a 3-clique community, although the same cannot be said for Motif 3D2B.

We note that the first two considerations discussed in Sec.~\ref{subsec:challenges} necessitate careful implementation of the motif identification step in RGC (step 2 in Sec.~\ref{subsec:2drgc}). Specifically, we construct the contact network as an attributed \emph{multigraph} representation, wherein multiple contacts between rigid bodies are represented as multiple edges---this is the same representation as the contact network for body bar systems, but we note that in our case, the network changes as the number of nodes shrinks through the compression process (starting with a network in which all nodes are rods and no multi-edges are present). Through the course of the compression process, we preserve a mapping of the edges in the compressed network to the corresponding edges in the original rod contact network, in order to keep track of which contacts are rod-sharing. Furthermore, a node attribute is used to specify nodes as being singleton rods or non-axisymmetric rigid bodies. Finally, we address the third consideration of Sec.~\ref{subsec:challenges} (the importance of motif ordering) through randomizing the order in which the motifs are applied. In particular, we show that while different orderings of the motif identification/compression may lead to different identifications of rigid components (Appendix~\ref{app:compression}), this quandary appears to have little effect on the identification of the rigidity percolation threshold (Sec.~\ref{subsec:reshuffling}).

%% file: numerics.tex
\section{Numerical Experiments}
\label{sec:exp}
Like rod-hinge/socket systems in two dimensions, rod-socket systems in three dimensions undergo a rigidity percolation transition at a critical rod number density. Our experiments demonstrate that the current implementation of \emph{3D-RGC} appears to be an accurate means to characterize this transition. As there is no established method for exact rigidity characterization in such systems, we cannot support this claim using direct comparison as in \cite{rigidity_heroy}---rather, we show that rigid motifs not incorporated into this RGC implementation are very unlikely to occur in random simulations, thus giving the implementation credence (Sec.~\ref{subsec:accuracy}).

\begin{figure}[htb!]
\begin{center}
\includegraphics[width=\linewidth]{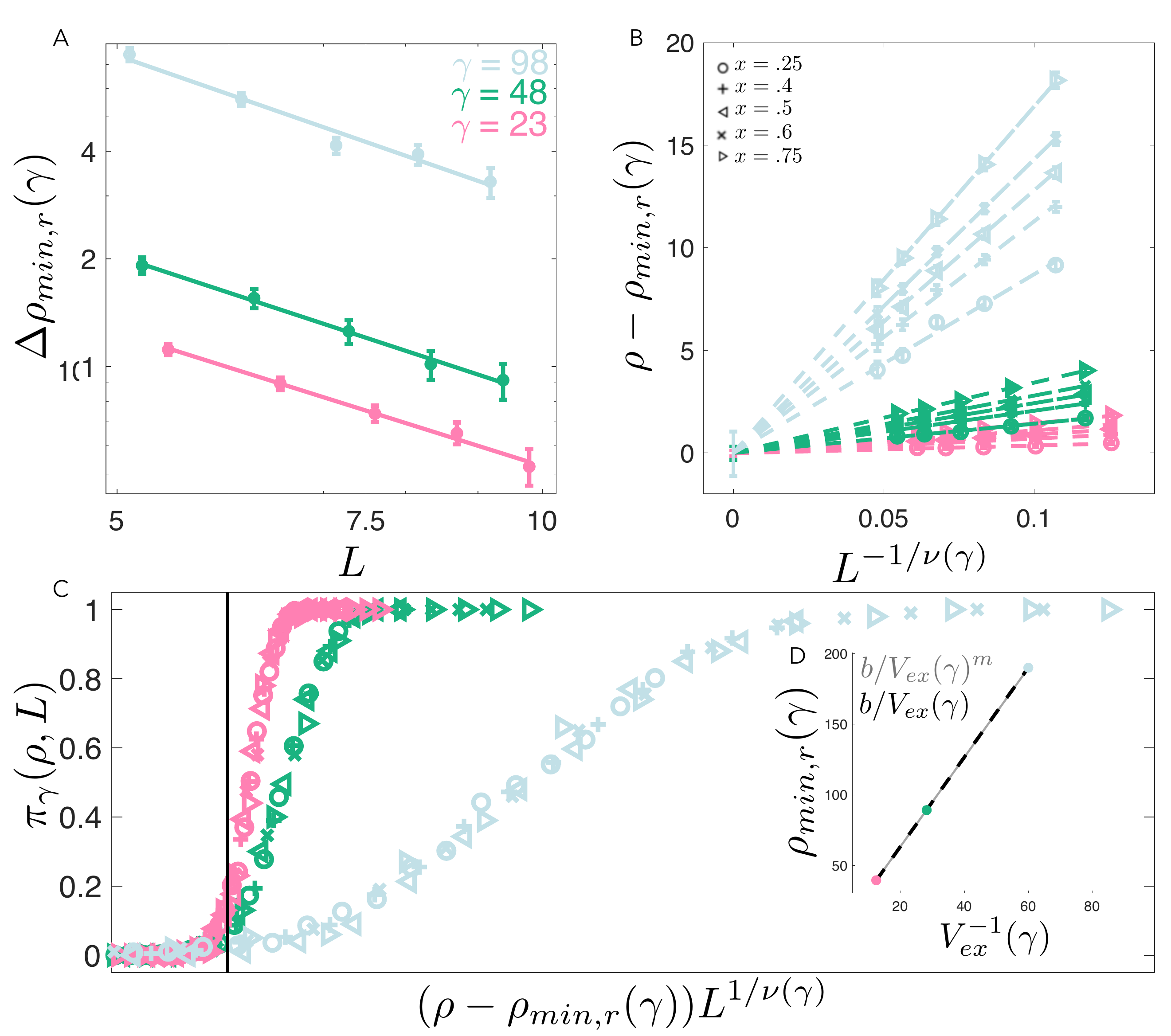}
\caption{\textbf{Estimation of parameters associated with the rigidity percolation transition for varying rod aspect ratio.} A. We calculate $\Delta \rho_{min,r}(\gamma)$ via the variation in $\pi_\gamma(\rho,\gamma)$ as described in Sec.~\ref{sec:fsa}, and then use Eq.~\ref{eq:delta} to find a robust estimate of $\nu(\gamma)$ for $\gamma\in\{23,48,98\}$. B. Then, we estimate the cumulative probabilities $\rho_x(L,\gamma)$, where $\pi_\gamma(\rho_x(L))>x$ for varying $x$, and fit these as linear functions of the $-1/\nu(\gamma)$-scaled system length $L$ to find a unified estimation of the transition point as $L\to\infty$ (each set of symbols corresponds to a box size). C. We then plot MC estimations of $\pi_\gamma(\rho,L)$ across all sampled parameters according to the scaling collapse ansatz given by  Eq.~\ref{eq:ansatz} and find this ansatz to be appropriate. Interestingly, the steepness of the finite-size scaling collapse decreases as $\gamma$ increases for fixed $L$. D. We find an inverse relationship between the rigidity percolation threshold and the average excluded volume of a rod ($V_ex$). The grey line denotes the linear regression fit wherein the exponent $m$ (slope in the log-log plot) is allowed to vary, while the black dotted line has a fixed exponent of 1 (only the intercept varies).}
\label{fig:scaling}
\end{center}
\end{figure}

\subsection{Experimental protocol}
\label{subsec:exp}
In order to study the rigidity percolation transition, we implement \emph{3D-RGC} on systems of varying rod number density, wherein the rods are monodisperse spherocylinders with aspect ratio $\gamma=\frac{\ell}{2r} \in\{23,48,98\}$, which are randomly placed and isotropically oriented in cubic boxes (with periodic boundary conditions) of varying system length $5<L<10$. For the remainder of this paper, we assume $L$ has been nondimensionalized by $\ell$ (i.e., $L/\ell \to L$). We perform $100$--$500$ Monte Carlo (MC) simulations at each nondimensionalized rod number density $\rho=N_R/L^3$ ($N_R$ is the number of rods), and we run more simulations for smaller $L$, since we find that their associated rigidity transitions are steeper than those associated with larger $L$. For each simulation, we use \emph{3D-RGC} to identify the absence/presence of a spanning rigid cluster that contacts both system boundaries along one dimension. Since \emph{3D-RGC} identifies no false positives, we identify a system as having a spanning rigid component if any of the trials are positive. Finally, we perform the same analyses at lower rod densities in order to compute the contact percolation threshold, which we associate the presence of a component that contacts both system boundaries along one dimension.

\subsection{Finite-size scaling analysis}
\label{sec:fsa}
In accord with percolation theory \cite{stauffer2018introduction}, we take the ansatz that the probability $\pi_\gamma(\rho,L)$ of a system containing a spanning rigid component varies \emph{for fixed $\gamma$} as the difference between the system's particle density and the rigidity percolation threshold $\rho_{min,r}(\gamma)$. This difference is scaled by some inverse power of the system length. This power is called the correlation length exponent $\nu(\gamma)$, and is associated with the divergence of the probability of a system having a spanning component about $\rho_{min,r}(\gamma)$:
\begin{equation}
\pi_\gamma(\rho,L)=f_\gamma\left([\rho-\rho_{min,r}(\gamma)]L^{1/\nu(\gamma)}\right).
\label{eq:ansatz}
\end{equation}
We expand this equation for $\rho\rightarrow \rho_{min,r}(\gamma)$ to give 
\begin{equation}
\frac{d\pi_\gamma}{d\rho}=L^{1/\nu(\gamma)}\frac{df_\gamma}{d\rho}\left([\rho-\rho_{min,r}(\gamma)]L^{1/\nu(\gamma)}\right)\,,
\end{equation}
which we invert to find 
\begin{equation}
\Delta \rho_{min,r}(L,\gamma)=f_\gamma L^{-1/\nu(\gamma)},
\label{eq:delta}
\end{equation}
where $\Delta\rho_{min,r}(L,\gamma)$ is the deviation of the rigidity percolation threshold at fixed $L$, which may be defined (though we do not actually compute it in this way, favoring the method of the next paragraph as in \cite{jacobs1996generic,rigidity_fibers2d,rigidity_heroy}) as $\langle \sqrt{(\rho_{est}-\rho_{av})^2}\rangle.$ Here, $\rho_{est}$ is the density at which a spanning rigid cluster first appears for a particular set of simulations across sampled $\rho$ values at fixed $(L,\gamma)$ and $\rho_{av}$ is the average across all sets of simulations. Finally, $f_\gamma$ is a prefactor that we find has dependence on $\gamma$.

Our general procedure is to first estimate the correlation length exponent using Eq.~\ref{eq:delta}, and then calculate the transition point $\rho_{min,r}(\gamma)$. More specifically, we estimate $\pi_\gamma(\rho,L)$ as the Monte Carlo average probability of a system having a rigid spanning component for varying $(\rho,L)$ at each $\gamma$. Then, we estimate $\Delta \rho_{min,r}(L,\gamma)$ by first fitting for each $(L,\gamma)$ a cumulative logistic distribution $F_{L,\gamma}(\rho; \mu,s)=1/(1+e^{-\frac{\rho-\mu}{s}})$ to the observations $\pi_\gamma(\rho,L)$, and then setting $\Delta \rho_{min,r}(L,\gamma)$ equal to the standard deviation $s\pi/\sqrt{3}$ about the estimated mean $\mu$. (Parameter $s$ is a scale parameter proportional to the standard deviation.) Then, we perform linear regression of $\log (\Delta \rho_{min,r}(L,\gamma))$ against $\log L$ to estimate $\nu(\gamma)$ via Eq.~\ref{eq:delta} (see Fig.~\ref{fig:scaling}). We use a simple resampling (with replacement) method to simultaneously determine confidence intervals for $\nu(\gamma)$ by resampling estimates of $\tilde{\pi}_\gamma(\rho,L)$ and using the same procedure as above to estimate $\tilde{\nu}(\gamma)$. We use the tilde to indicate one resampling across all $(\rho,L,\gamma)$. We then use the collection of these resampled estimates to calculate the corresponding $95 \%$ confidence intervals for the critical parameters, which are presented along with the corresponding estimates for each $\gamma$ in Table~\ref{tab:1}.
\begin{table}[h]
\centering
    \begin{subtable}{\textwidth}
        \centering
        \caption{Estimates (and $95\%$ confidence intervals) of the correlation length exponent and the rigidity percolation threshold.}
        \begin{tabular}{c|c|c|c}
	&  $\gamma=23$ &  $\gamma=48$ & $\gamma=98$\\
	\hline
	$\rho_{min,r}(\gamma)$ & 39.51 (39.35, 39.64) &89.18 (88.82, 89.49) &190.00 (188.92, 190.98)\\
	$\nu(\gamma)$  &0.804 (.705, .920)  &0.769 (.673, .891)&0.729 (.655, .806)
	\end{tabular}
        \label{tab:1}
    \end{subtable}
    \\
     \hfill
     \begin{subtable}{\textwidth}
       \centering
       \caption{Estimates (and $95\%$ confidence intervals) of rod contact numbers at the rigidity percolation threshold.}
       \begin{tabular}{c|c|c|c}
       Contact number &  $\gamma=23$ &  $\gamma=48$ & $\gamma=98$\\
       \hline
      $\langle  c_0(\rho_{min,r}(\gamma))\rangle$ & 3.18 (3.17, 3.19)&3.17 (3.15, 3.18)&3.17 (3.15, 3.19)\\
      $\langle  c_1(\rho_{min,r}(\gamma))\rangle$ & 3.35 (3.34, 3.36)&3.33 (3.32, 3.34)&3.33 (3.32, 3.35)\\
      $\langle  c_2(\rho_{min,r}(\gamma))\rangle$ & 3.61 (3.60, 3.61)&3.58 (3.57, 3.59)&3.58 (3.57, 3.59)
       \end{tabular}
       \label{tab:3}
       \end{subtable}
    \\
    \hfill
    \begin{subtable}{\textwidth}
        \centering
        \caption{Estimates (and $95\%$ confidence intervals) for the contact percolation transition points, the associated correlation length exponents, and the rod contact numbers at the respective thresholds.}
        \begin{tabular}{c|c|c|c}
        &  $\gamma=23$ &  $\gamma=48$ & $\gamma=98$\\
       \hline
       $\rho_{min,c}(\gamma)$ &19.21 (19.14,19.27) & 38.69(38.55,38.83)&75.10(74.89,75.33)\\
       $\nu^c(\gamma)$ &0.847(.781,.911) &0.841(.775,.914) &0.801(.729,.879) \\
      $\langle  c_0(\rho_{min,r}(\gamma))\rangle$ & 1.55 (1.54,1.55)& 1.37(1.37,1.38)&1.25(1.25,1.26)
      
       \end{tabular}
       \label{tab:2}
       \end{subtable}
     
\label{tab:temps}
\end{table}

Having estimated $\nu(\gamma)$, we now expand Eq.~\ref{eq:ansatz} around $\rho=\rho_{min,r}(\gamma)$ and invert, deriving the condition that
\begin{equation}
\rho_x(L,\gamma)=(\mathrm{constant})\cdot L^{-1/\nu(\gamma)}+\rho_{min,r}(\gamma),
\label{eq:extrap}
\end{equation}
where $\rho_{x}(L,\gamma)$ is the probability distribution such that $\pi_\gamma(\rho_x(L,\gamma),L)=x$ for some $x\in[0,1]$. We use this equation to extrapolate the values of $\rho_x(L,\gamma)$ as $L\to\infty$  for $x\in\{0.25,0.4,0.5,0.6,0.75\}$, which allows us to estimate $\rho_{min,r}(\gamma)$ using Eq.~\ref{eq:extrap} and the following procedure. First, we estimate $\rho_x(L,\gamma)$ for each $x$ and $L$ via inverse prediction from the corresponding cumulative distributions $F_{L,\gamma}(\rho;\mu,s)$ that we have already fitted. Then, we use least squares linear regression to fit $\rho_x(L,\gamma)$ against $L^{-1/\nu(\gamma)}$ for each $x$, given the constraint\footnote{When we relax this assumption, each of the individual fits (for varying $L$) have an intercept that is within $.01$ rods per unit volume of the pooled estimate.} that the collection of these fitted lines must intersect as $L\to\infty$ in accord with the hypothesis that $\rho_{x,\gamma}(\infty)$ is constant for $x\in(0,1)$. We thereby estimate $\rho_{min,r}(\gamma)$ and find the scaling collapse ansatz according to Eq.~\ref{eq:ansatz} to be accurate for each $\gamma$ (see Fig.~\ref{fig:scaling} and Table~\ref{tab:1}). We again use a case resampling procedure to determine confidence intervals for $\rho_{min,r}(\gamma)$---we use the same estimated sample values $\{\tilde{\nu}(\gamma)\}$ as above, and for each sample estimate $\tilde{\rho}_{min,r}(\gamma)$, we use Eq.~\ref{eq:extrap} again to fit each $\tilde{\rho}_{x,\gamma}(L)$ against $L^{-1/\tilde{\nu}(\gamma)}$ so as to simultaneously estimate confidence intervals for both $\nu(\gamma)$ and $\rho_{min,r}(\gamma)$. Finally, we repeat this analysis at lower densities to compute the contact percolation thresholds $\rho_{min,c}$. We present our results in Table~\ref{tab:2}.

\subsection{Critical contact numbers}
For each simulated rod-socket system in the previous section, we calculate the empirical mean contact number\footnote{The findings agree fairly well with the rod contact equation derived in \cite{philipse1996random}.} (or mean degree) $\langle c_0\rangle \doteq \frac{2N_c}{N_R}$ as well as the contact numbers within the 1- and 2-core ($\langle c_1\rangle, \langle c_2 \rangle$).\footnote{The 1-core corresponds to the largest connected component of the rod contact graph and the 2-core corresponds to the graph's largest connected subgraph in which the minimum degree is 2.} Given MC estimates of these contact numbers for varying $(\rho,L,\gamma)$, we then calculate linear fits of each against $\rho$ for each $(L,\gamma)$ (we note the fits are very strong: $R^2>.97$) and predict the critical contact numbers that correspond to the estimated $\rho_{min,r}(\gamma)$ as well as the corresponding confidence intervals\footnote{We calculate these by simply using the same linear fit of the contact numbers against $\rho$ to linearly predict the contact numbers at each end of the $95\%$ confidence intervals for $\rho_{min,r}(\gamma)$. As noted, the linear fits are quite precise: the uncertainty of the linear predictions is generally smaller than the uncertainty of the $\rho_{min,r}(L,\gamma)$ estimations.} (see Table~\ref{tab:3}).

\begin{figure}[htbp]
\begin{center}
\includegraphics[width=.9\linewidth]{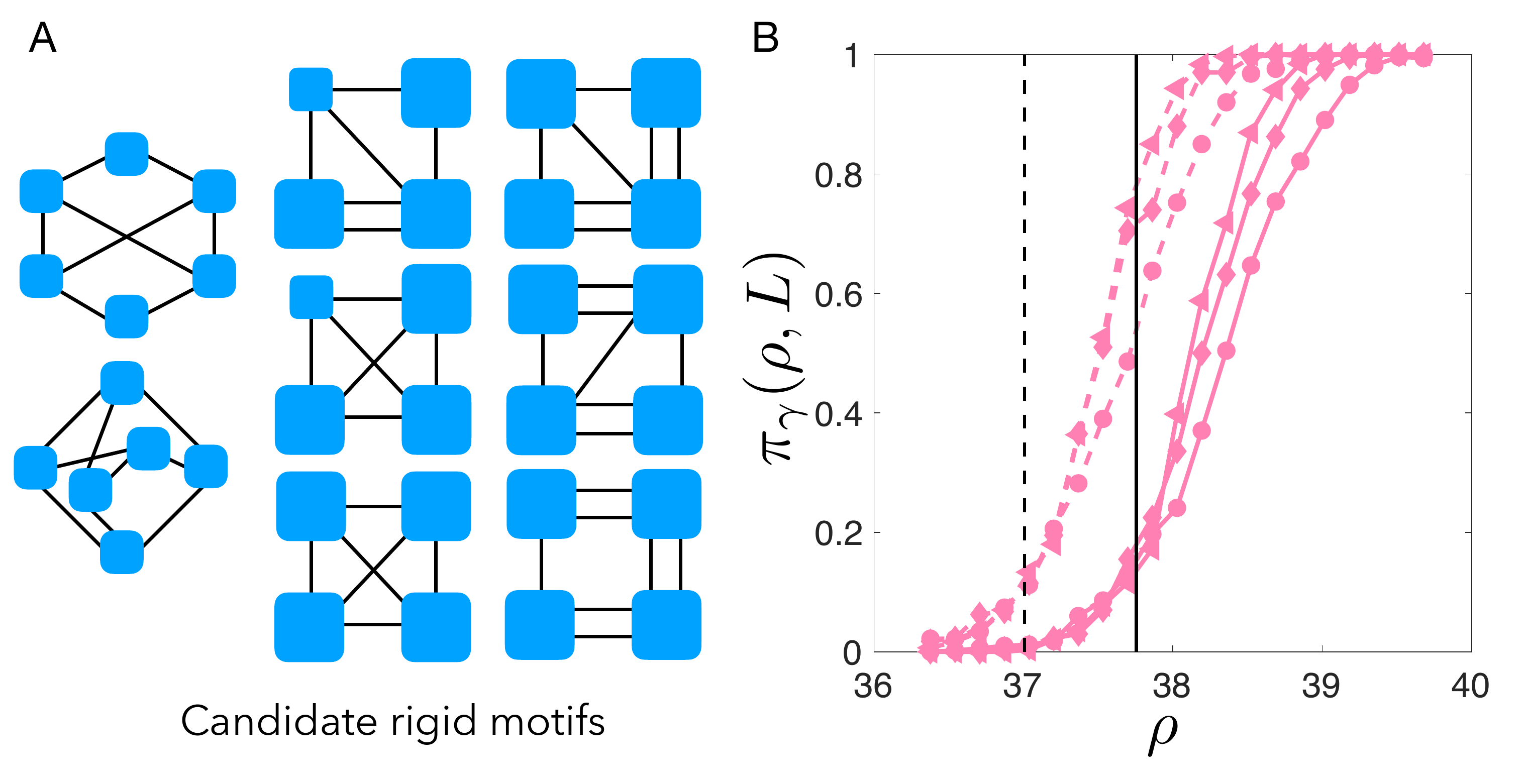}
\caption{\textbf{Candidate rigid motifs and possible effects on the rigidity percolation threshold.} A: We identify 6 possibly rigid  motifs of size $N_r=4$ containing either rods/nonaxisymmetric rigid bodies as the included particles, and 2 rigid motifs of size $N_r=6$ containing only rods as the included particles. B. Including all these candidate rigid motifs into \emph{3D-RGC} lowers the rigidity percolation threshold (dashed vertical line) by $3.9\%$. We conclude from this analysis that the threshold we identify in the previous subsections (vertical solid line) is likely an overestimate of the true threshold (for each $\gamma$), but that the threshold is not likely to be very much lower.}
\label{fig:accuracy}
\end{center}
\end{figure}

\subsection{Global consistency}
\label{subsec:reshuffling}
As discussed in Appendix~\ref{app:compression}, \emph{3D-RGC} may identify different rigid components in different implementations on the same rod-socket system, as the final results depend on the order in which rigid motifs are identified and compressed. Determining whether or not a system contains a spanning rigid component may therefore require an exhaustive number of stochastic \emph{3D-RGC} implementations. Because the algorithm is sufficient but not necessary, any single positive implementation (detecting a spanning rigid component) indicates that a system has a spanning component. However, we find that in practice, there are relatively few discrepancies across implementations and that the difference between running one implementation and running 40 is rarely significant (see Fig.~\ref{fig:accuracy}A). Moreover, the estimates of $\nu(\gamma)$ and $\rho_{min,r}(\gamma)$ have little sensitivity to the number of implementations used. Using a single implementation results in less than $0.001\%$ relative difference in the estimation of $\rho_{min,r}(\gamma)$ as compared to using the union of $40$ implementations. We performed the procedure of Sec.~\ref{sec:fsa} for random subsets of these $40$ implementations at $\gamma=48$ and found that using $10$ implementations almost always ($>90\%$ of the time) results in the \emph{exact} same estimations of $\rho_{min,r}(\gamma)$ and $\nu(\gamma)$ as using $40.$ Because we're most interested in capturing the nature of the transition across $\gamma$ and in relation to $\rho_{min,c}(\gamma)$ (alongside computational burdens) rather than pinpointing the transition point to the absolutely highest precision possible, we only use 10 implementations for the calculations of the previous two subsections.

\subsection{Accuracy}
\label{subsec:accuracy}
In Sec.~\ref{subsec:3dmotifs} and Appendix~\ref{app:rigidity proofs}, we rigorously prove that the rigid motifs incorporated into \emph{3D-RGC} are indeed rigid. Any rod-socket system identified as rigid using this algorithm is therefore rigid. However, it is not the case that every rigid rod-socket system would be identified as rigid using this algorithm---i.e., the algorithm is \emph{sufficient but not necessary}. On the other hand, we can identify several \emph{necessary but insufficient} conditions for rigidity. In particular, any rigid motif (wherein the individual components are rods) must identify Maxwell's isostatic condition, as there must be sufficient constraints to bind all degrees of freedom. Secondly, the rigid motif must be contained within its two-core (any rods with less than two contacts would necessarily have nontrivial degrees of freedom).  Finally, a motif must contain sufficient contacts in its rigidity matrix in order to constrain the inherent particles' degrees of freedom (this condition encompasses the former two).

In order to assess the accuracy of \emph{3D-RGC} at a local scale (i.e.\ to address whether we have identified a sufficient number of rigid motifs to design an accurate algorithm), we examine a database containing all possible respectively non-isomorphic, connected graphs of small sizes \cite{graphlet}---and compare the results of \emph{3D-RGC} with implementation of these other conditions on these graphs. Specifically, for graphs containing $n_r$ nodes (rods), we implement $\emph{3D-RGC}$ and also examine where they meet the necessary but insufficient conditions for rigidity discussed above. If they do meet such conditions, we label the graphs as ``candidate rigid motifs.'' We find that \emph{3D-RGC} identifies as rigid all candidate rigid motifs of size $n_r\le 5$ (see Fig.~\ref{fig:accuracy}), but does not identify $2$ potentially rigid motifs of size $n_r=6$. The only candidate motifs of size $N_r=7$ that the algorithm does not identify contain one of these $n_r=6$ candidate motifs as subgraphs. However, we also detect 8 irreducible candidate motifs of size $N_r=8$, suggesting it is unfeasible to count all rigid motifs.

We hypothesize, in supposing \emph{3D-RGC} to be an accurate approach, that these other candidate motifs and larger ones that we have not detected are exceedingly rare in random systems. However, we also use the same analysis wherein all the rigid components are allowed to be non-axisymmetric rigid bodies, and find that \emph{3D-RGC} identifies all rigid bodies with $\le 3$ components in this case but does not identify 6 (irreducible) motifs containing $4$ components. Inclusion of these multi-scale motifs is, in our intuition, more likely to influence the rigidity decomposition.

Finally, we use these candidate motifs to characterize \emph{3D-RGC}'s accuracy for estimation of the rigidity percolation threshold. We implement a second version of \emph{3D-RGC} wherein we include the candidate rigid motifs of size $N_r\le 7$ (including higher-scale motifs described above) (see Fig.~\ref{fig:accuracy}) in addition to the motifs included in our simpler base version of \emph{3D-RGC} used above. We examine the output of this algorithm for the same candidate $\gamma=23$, lower $L$ systems as above (the more inclusive second version becomes prohibitively slow on larger systems). Using this approach, we repeat the procedures of Sec.~\ref{sec:fsa} in order to estimate a new rigidity percolation threshold of $\rho_{min,r}=38.01$ ($95\%$ CI: $37.75,38.26$). We cannot conclusively state whether this newly identified threshold is either above/below the true threshold, as the candidate motifs we identify in this subsection are only possibly rigid (they contain the appropriate number of constraints in their rigidity matroid). However, the relative closeness of this new estimate to that we identify in Sec.~\ref{sec:fsa} ($\rho_{min,r}=39.51$) indicates that sequentially adding additional motifs will likely have only a tempered effect on the resulting estimate. Adding more motifs will tend to monotonically decrease the threshold, but the effect size decays with the complexity of the candidate motifs. Therefore, we conclude that the estimated rigidity percolation thresholds identified in Sec.~\ref{sec:fsa} are very close to the true thresholds, and that our insights regarding the nature of the rigidity transition are valid.

%% file: analyses.tex
\begin{figure}
    \centering
    \includegraphics[width=\linewidth]{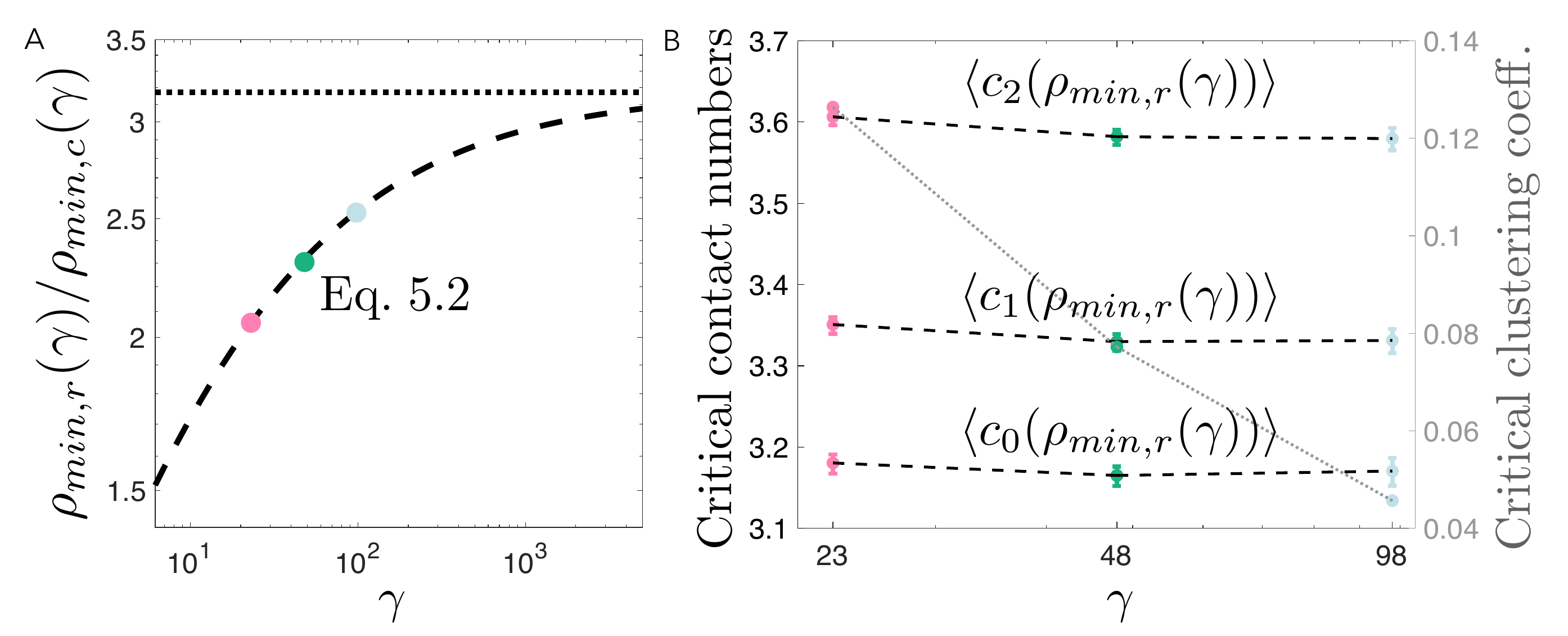}
    \caption{{\bf{The relationship between  and contact percolation thresholds and the Maxwell prediction.}} A. The ratio of the identified rigidity percolation threshold to its contact analogue increases with aspect ratio and seems to converge near $\pi$ in the slender body limit. The markers indicate our numerical findings, while the line is the prediction of Eq.~\ref{eq:ratio}. B. We find a nearly constant relationship between the critical contact numbers (mean degree, and 1-core and 2-core mean degrees at the estimated rigidity percolation threshold) across $\gamma$, even as higher aspect ratio rod networks have different network properties. In particular, the dotted line in panel B depicts the mean clustering coefficient \cite{newman2009random} for the different contact networks.}
    \label{fig:predictions}
\end{figure}

\section{Interpretation of numerical findings}
\label{sec:interp}
We now examine the implications of our findings from Sec.~\ref{sec:fsa} in more detail. In particular, we will more closely consider the following: (i) the relationship between excluded volume and the rigidity percolation transition; (ii) the agreement with our findings and Maxwell's isostatic condition for the critical rod contact number; and (iii) the ratio of the rigidity percolation critical density to the contact percolation critical density.
\subsection{Excluded volume and the contact percolation threshold}
\label{subsec:contact}
Here, we briefly review and reproduce certain findings with regards to contact percolation for the purpose of comparing the rigidity percolation transition to the contact percolation transition. Specifically, we note a finding that was originally proposed by \cite{balberg1984percolation} and analytically derived by \cite{bug1985continuum} applying to systems as ours: the contact percolation transition density has reciprocal dependence on the average excluded volume of a rod, $V_{ex}(\gamma)$, in the slender body limit. The excluded volume of a rod refers to the volume of space containing the rod in which no other rod can be centered if the two rods are not overlapping. More recently, it has been shown that this slender body limit is approached rather slowly, and that the constant has a power law dependence on the rod aspect ratio \cite{neda1999reconsideration}:
\begin{equation}
    \rho_{min,c}=\frac{1+s(\gamma)}{V_{ex}(\gamma)},
\label{eq:excl}
\end{equation}
where $s(\gamma)\to 0$ as $\gamma\to\infty$. When we combine Eq.~\ref{eq:excl} with the rod contact equation derived by Phillipse \emph{et al.} \cite{philipse1996random}, we find that at the percolation transition, the contact number $\langle c_0(\rho_{min,c}(\gamma))\rangle=1+s(\gamma)$, which is a simple and interesting result shown by \cite{balberg1984percolation}. This finding --- which we confirm experimentally in Fig.~\ref{fig:contact} in the Appendices --- can also be compared to the percolation transition for Erd\H{o}s-R\'enyi networks. In Erd\H{o}s-R\'enyi networks, which are characterized by nodes having equivalent binomial degree distributions, the contact percolation transition coincides with the contact number surpassing one (that is, in the large network limit) \cite{erdHos1960evolution}.

\subsection{Rigidity percolation and a proposition for the  slender body limit of $\rho_{r,min}/\rho_{c,min}$}
We observe in Fig.~\ref{fig:scaling} an apparent reciprocal dependence of $\rho_{min,r}(\gamma)$ on $V_{ex(\gamma)}$, which is similar to what we observe for the contact percolation case. However, for the aspect ratios that we studied, we do not find a parameter that depends on the slender body approximation as $s(\gamma)$ in Eq.~\ref{eq:excl}. Instead, we find evidence for a constant multiplicative factor $b$  that is valid across all $\gamma,$
\begin{equation}
\rho_{min,r}(\gamma)=b/V_{ex}(\gamma).
\label{eq:predexc}
\end{equation}
Our ability to determine the precise nature of the relationship between $\rho_{min,r}(\gamma)$ and $V_{ex}(\gamma)$ is limited (as we only have three data points), but we find evidence nonetheless that this simple model is a good fit. Letting $\mathcal{P}_\gamma$ be the random variable for the bootstrapped distribution of $\rho_{min,r}$, we first fit (using linear regression) for each sample in the distribution the equation $\mathcal{P}_\gamma=m/V_{ex}(\gamma)+\beta$. We then use these bootstrapped fits to calculate $95\%$ confidence intervals for $m\in[0.994,1.002]$ and find that we cannot reject the null hypothesis that $m=1.$ This suggests that $\rho_{min,r}$ varies exactly with $V_{ex(\gamma)}^{-1}$ as in the contact percolation case. We display the mean of the two-parameter fit as well as $95\%$ confidence intervals for the fit in Fig.~\ref{fig:scaling}D. Additionally, we display the naive fit of $\rho_{min,r}(\gamma)=b/V_{ex}(\gamma)$ (wherein $b=3.17$ is simply the mean value of $\rho_{min,r}(\gamma)\times V_{ex}(\gamma)$).

Now, we pay heed to the implications of Eq.~\ref{eq:predexc}. Combining this with Eq.~\ref{eq:excl}, we observe that 
\begin{equation}
    \frac{\rho_{min,r}(\gamma)}{\rho_{min,c}(\gamma)}=\frac{b}{1+s(\gamma)}
\label{eq:ratio}
\end{equation}
If we take the slender body assumption ($r\to 0$), then $s\to 0$ and the ratio converges to $b=3.17$ (Fig.~\ref{fig:predictions}A). Regarding the sensitivity of this fit to approximate thresholds identified from \emph{3D-RGC}, more work is required but we might surmise that only $b$ in Eq.~\ref{eq:predexc} is sensitive to the algorithm's accuracy. If this conjecture holds, then the ratio in Eq,~\ref{eq:ratio} will vary as $1/(1+s(\gamma))$ but the fixed point will decrease as the approximate threshold approaches the true one.

\subsection{Maxwell's isostatic condition}
Now, we assess the accuracy of Maxwell's isostatic condition in this system. Using the rod-contact equation, given by $\rho=\langle c \rangle/V_{ex}$ \cite{philipse1996random}, we can use the prediction of Eq.~\ref{eq:predexc} to find that 
\begin{equation}
\langle c_0(\rho_{min,r}(\gamma)\rangle=b,
\label{eq:contact_pred}
\end{equation}
which --- like the prediction of Maxwell's isostatic condition (Sec.~\ref{sec:intro}) --- is independent of $\gamma$, but is lower than Maxwell's estimate $\langle c_0(\rho_{min,r}(\gamma)\rangle=10/3$.

Empirically (Fig.~\ref{fig:predictions}B), we find that the critical contact number $\langle  c_0(\rho_{min,r}(\gamma))\rangle$ agrees well with Eq.~\ref{eq:contact_pred}. However, we find that the mean contact number within the giant component $\langle  c_1(\rho_{min,r}(\gamma))\rangle$ approximately coincides with the prediction using Maxwell's isostatic condition. This discrepancy indicates that at the rigidity percolation transition point, many rods may still be disconnected even as a critical mass forms the giant rigid component. Moreover, imposing the constraint that all rods have degree 2---which is a prerequisite for rigidity---increases the critical contact number $\langle  c_2(\rho_{min,r}(\gamma))\rangle$ far beyond the Maxwell estimate. This finding demonstrates that a nontrivial number of redundant constraints are present in these random systems at the rigidity percolation threshold. Finally, because \emph{3D-RGC} represents a sufficient but not necessary characterization of rigidity percolation, these critical contact numbers all represent upper bounds on the true critical contact numbers. The Maxwell prediction is therefore possibly a useful summary approximation, but overestimates the true threshold.

While the critical contact numbers are at least similar for varying aspect ratios, we find for fixed $L$ that $\Delta \rho_{min,r}$ increases with $\gamma$ (Fig.~\ref{fig:scaling}A). This dependence coincides with a a decreasing relationship between rod aspect ratio and the average clustering coefficient of the network (Fig.~\ref{fig:predictions}B), which measures the average tendency of rods to participate in triangular motifs \cite{newman2009random}. Based on these findings, we hypothesize that, for fixed system size, rod clustering is associated with the width of the range over which $0<\pi_\gamma(L,\gamma)<1$. While systems of equal rod density but varying aspect ratio are expected to have equal contact numbers, lower aspect ratio systems have more rigid motifs (especially Motif 3D3A, which has a prevalence that is directly proportional to the clustering coefficient). This tendency has no effect on $b/V_{ex}(\gamma)$, but seems to have a decreasing effect on $\Delta \rho_{min,r}$.
 
\subsection{Possible aspect ratio invariance of the correlation length exponent}
Our ability to infer the correlation length exponent is especially limited because we cannot bound it (as we can $\rho_{min,r}$), but our findings suggest that it is likely to be similar across $\gamma$. Generally, we find similar estimates ($0.7$--$0.8$) for the correlation length exponent $\nu(\gamma)$ across the three rod aspect ratios that we studied (see Table~\ref{tab:3}). Because we only choose 5 system sizes for each aspect ratio due to computational limitations, the precision of our estimation of $\nu(\gamma)$ is limited. Treat each sampling distribution of $\nu(\gamma)$ as an instance of a random variable $\mathcal{N}_\gamma$, we compare the distributions of these random variables. We find only marginally significant evidence ($p=.09$) for inequality - in particular, that $\mathcal{N}_{50}$ has a lower mean than both $\mathcal{N}_{25}$ and $\mathcal{N}_{12.5}$.

%% file: discussion.tex
\section{Discussion}
\label{sec:discussion}
It is well known for completely penetrable or `soft-core' isotropically oriented rods of aspect ratio $\gamma$ that the contact percolation threshold varies with the reciprocal of the excluded volume of a rod in the slender body limit. The slender body limit, which can be derived analytically, is approached rather slowly in practice ($s>0.2$ for $\gamma<100$) \cite{philipse1996random,neda1999reconsideration,bug1985continuum}. This dependence on excluded volume arises in Eq.~\ref{eq:excl} and has two important corollaries for high aspect ratio rods: (1) the critical contact number at the contact percolation transition point is equal to $1+s(\gamma)$, implying that it is only approximately invariant for very slender rods ($\gamma\gg 100$); and (2) the transition packing fraction $\phi_{min,c}(\gamma)=\rho_{min,c}(\gamma)\times\left(\frac{\pi}{4\gamma^2} + \frac{\pi}{6\gamma^3}\right)$\footnote{The right hand factor here is simply the volume of a rod divided by $\ell^3$, which comes about due to nondimensionalization of the number density.} scales approximately linearly with the rod aspect ratio $\gamma$ for high aspect ratio rods \cite{philipse1996random}. We observe here that rigidity percolation also varies with the reciprocal of the excluded volume of a rod, but the slender body approximation is accurate at a much lower aspect ratio. This means that both (1) and (2) are true for rigidity percolation, and the caveat regarding sufficiently high aspect ratio rods is unnecessary (at least for $\gamma\ge 23$).

Why would the rigidity percolation transition achieve high aspect ratio behavior at a lower aspect ratio than the contact analogue? This question merits further study, but we suspect that the origin of the deviation for lower aspect ratios may lie in network properties. For instance, low-aspect-ratio rod systems have highly clustered contacts (see the dependence of the average rod clustering coefficient on aspect ratio in the rod contact networks, Fig.~\ref{fig:predictions}) that differentiate these networks from random graphs. Chatterjee and Grimaldi \cite{chatterjee2015random} formulate these networks as random geographic networks to derive the contact percolation threshold from degree distributions. Perhaps this approach might be extended to recover the $s(\gamma)$ dependence on aspect ratio. Better understanding of this contact percolation aspect ratio deviation might help us improve our understanding of the lack of such deviation for rigidity percolation.

Our investigations here are motivated by a desire to better understand the nature of the rheological percolation threshold that is observed in a variety of composite materials \cite{favier1997mechanical,hough2004viscoelasticity,celzard,penu}. One overlooked but easily observed feature of many of these composites is the spatial heterogeneity of the stiffening phase. For instance, carbon nanotubes and other nanoparticles often tend to be highly clustered, leading to departure from the behaviors expected in systems wherein particles have uniformly random position. Additionally, many composites may experience degrees of alignment on account of preparation stages and complex physical phenomena. We surmise that this heterogeneity gives rise to the wide variety of experimentally calculated ratios of the rheological percolation threshold to the conductivity threshold, which has been observed to be as high as 3 and as low as 1 \cite{penu}. Whereas the glass forming condition predicts this ratio (for uniformly random systems) to be $1.5$ and another mean field prediction based on semiflexible fiber system predicts it to be $5/3$ (see Sec.~\ref{sec:intro}), we estimate the ratio to have aspect ratio dependence and approach $b=3.17$ for slender body rods. Future work connecting contact/rigidity percolation to relevant experimental findings might examine the dependence of the corresponding thresholds upon spatial heterogeneity within the rod network, since agglomeration (see \cite{silva} for an example) might lead to drastically different estimates from the uniformly random case. Additionally, we might expect that the nature of interactions between `hard-core' (impenetrable) rods \cite{schilling2015percolation,berhan2007modeling,chatterjee2015random} will affect the position of the rigidity percolation threshold, since it also has also been shown to affect the contact percolation threshold.

Finally, future work should also aim to connect the current findings to other studies in rigidity percolation in order to understand how the network generation (e.g.\ random rods vs.\ diluted lattices) influences the associated transition. Latva-Kokko and Timonen show that rigidity percolation on 2D rod-hinge networks (`Mikado networks') fall within the same universality class as rigidity percolation in 2D central-force networks, showing agreement between the respective correlation length exponents and fractal dimensions of the incipient giant clusters \cite{rigidity_fibers2d}. However, rigidity percolation and contact percolation in these networks fall under different classes (see \cite{chaikin1995principles} for a thorough discussion of universality classes). The rigidity percolation transition in 3D diluted central-force networks is thought to be first-order (i.e., it is discontinuous in an appropriate order parameter) while that for 3D bond-bending networks is second-order (i.e., the order parameter varies continuously, but its derivative is discontinuous) \cite{chubynsky2007algorithms}. There is not sufficient evidence here to definitively conclude the order of the transition in our system (though visual examination of the scaling collapse in Fig.~\ref{fig:scaling} tempts us to surmise that it is continuous, we have not as yet conducted any rigorous tests), nor to attempt to place it in a known universality class. It would be interesting in future work to explore whether our RGC framework can aid in establishing a rigorous connection between rigidity matroid theory and universality classes.

%% file: proofs.tex
Here, we present and sketch the proofs of many rigid motifs for the construction of composite rigid components in three dimensions. As was the case in two dimensions \cite{rigidity_heroy}, this list is nonexhaustive (and there is no reason to suggest constructing an exhaustive list is possible). In order to avoid creating a new litany of symbols, we do not distinguish rods from other rigid components --- both are called $R_i$ for some $i$ --- nor do we notationally distinguish rod-sharing contacts from other contacts. Rather, we just indicate the number of rods contained in rigid component $i$ as $n_r^i$. Finally, we note that in cases in which there is more than one motif with $x$ rigid bodies (as the distinction between rods from other rigid bodies opens up this possibility), we distinguish between the different rigid motifs by lettering: \emph{rigid motif 3DxA, 3DxB, \dots}.

\begin{theorem}
(Rigidity of Motif 3D3A): Let $R_1$, $R_2$, and $R_3$ be intersecting nonaxisymmetric rigid bodies ($n_r^1,n_r^2,n_r^3>1$) such that $R_1$ and $R_2$ intersect at $\ge 1$ instance, $R_1$ and $R_3$ intersect at $\ge2$ instances, and $R_2$ and $R_3$ intersect at $\ge2$ instances. Let $\bm p_1\in R_1\cap R_2$, let  $\bm p_2,\bm p_3$ be points in two of the instances comprising $R_1\cap R_3$, and let $\bm p_4, \bm p_5$ be points in two of the instances comprising $R_2\cap R_3$. The composition $R_1\cup R_2\cup R_3$ is rigid, unless one of the following occurs: $\bm p_1$ shares a rod with both $\bm p_2$ and $\bm p_3$; $\bm p_1$ shares a rod with both $\bm p_4$ and $\bm p_5$; or $\bm p_2$, $\bm p_3$, $\bm p_4$, and $\bm p_5$ all share the same rod (excluding these cases via hypothesis). 
\end{theorem}
\begin{proof}
First, we assume none of the rigid bodies are planar. We choose as the coordinate labelings $\{ \bm p_1, \bm p_2, \bm p_3, \bm p_{R_1}\}$ for $R_1$; $\{ \bm p_1, \bm p_4, \bm p_5, \bm p_{R_2}\}$ for $R_2$; and $\{ \bm p_2, \bm p_3, \bm p_4, \bm p_5\}$ for $R_3$, where $\bm p_{R_1}$ lies in $R_1 \backslash (R_2\cup R_3)$ and $\bm p_{R_2}$ lies in $R_2 \backslash (R_1\cup R_3)$. Combining the pairwise constraints within each labeling gives the rigidity matrix:
\begin{multline}
\bm X_1\ast \bm X_2\ast \bm X_3=
\tiny{\left[
\begin{array}{ccccccc}
\Delta \bm p_{1,R_1}&\bm 0&\bm 0&\bm 0&\bm 0&-\Delta \bm p_{1,R_1}&\bm 0\\
\bm 0 &\Delta \bm p_{2,R_1}&\bm 0&\bm 0&\bm 0&-\Delta \bm p_{2,R_1}&\bm 0\\
\bm 0 &\bm 0&\Delta \bm p_{3,R_1} &\bm 0&\bm 0& -\Delta \bm p_{3,R_1}&\bm 0\\
\Delta \bm p_{1,2} &-\Delta \bm p_{1,2} &\bm 0&\bm 0&\bm 0&\bm 0&\bm 0\\
\Delta \bm p_{1,3} &\bm 0 &-\Delta \bm p_{1,3} &\bm 0&\bm 0&\bm 0&\bm 0\\
\bm 0& \Delta \bm p_{2,3} &-\Delta \bm p_{2,3} &\bm 0&\bm 0&\bm 0&\bm 0\\
\Delta \bm p_{1,R_2} &\bm 0&\bm 0&\bm 0&\bm 0&\bm0&-\Delta \bm p_{1,R_2} \\
\bm 0 &\bm 0&\bm 0&\Delta \bm p_{4,R_2} &\bm 0&\bm 0&-\Delta \bm p_{4,R_2}  \\
\bm 0 &\bm 0&\bm 0 &\bm 0& \Delta \bm p_{5,R_2} & \bm 0&-\Delta \bm p_{5,R_2}  \\
\Delta \bm p_{1,4} &\bm 0 &\bm 0&-\Delta \bm p_{1,4} &\bm 0&\bm 0&\bm 0\\
\Delta \bm p_{1,5} &\bm 0 &\bm 0 &\bm 0&-\Delta \bm p_{1,5} &\bm 0&\bm 0\\
\bm0 &\bm 0 &\bm 0 & \Delta \bm p_{4,5} &-\Delta \bm p_{4,5} &\bm 0&\bm 0\\
\bm 0& \Delta \bm p_{2,4} &\bm 0& -\Delta \bm p_{2,4} &\bm 0&\bm 0&\bm 0\\
\bm 0& \Delta \bm p_{2,5} &\bm 0&\bm 0& -\Delta \bm p_{2,5}  &\bm 0&\bm 0\\
\bm 0& \bm0&\Delta \bm p_{3,4} &-\Delta \bm p_{3,4} &\bm 0& \bm 0 &\bm 0\\
\bm 0& \bm 0&\Delta \bm p_{3,5} &\bm 0& -\Delta \bm p_{3,5}  &\bm 0&\bm 0
\end{array}
\right]},
\label{3daxiom3a_1}
\end{multline}
where the constraint rows $1$--$6$ derive from $R_1\cup R_3$, $7-12$ derive from $R_2\cup R_3$, and $13$--$16$ derive from $R_3$ only. We use row permutations to find that $\bm X_1 \ast \bm X_2 \ast \bm X_3$ is rank equivalent to the block triangular matrix:
\begin{multline}
\scriptsize{
\left[
\begin{array}{ccccccc}
\Delta \bm p_{1,2} &-\Delta \bm p_{1,2} &\bm 0&\bm 0&\bm 0&\bm 0&\bm 0\\
\Delta \bm p_{1,3} &\bm 0 &-\Delta \bm p_{1,3} &\bm 0&\bm 0&\bm 0&\bm 0\\
\bm 0& \Delta \bm p_{2,3} &-\Delta \bm p_{2,3} &\bm 0&\bm 0&\bm 0&\bm 0\\
\Delta \bm p_{1,4} &\bm 0 &\bm 0&-\Delta \bm p_{1,4} &\bm 0&\bm 0&\bm 0\\
\bm 0& \Delta \bm p_{2,4} &\bm 0& -\Delta \bm p_{2,4} &\bm 0&\bm 0&\bm 0\\
\bm 0& \bm0&\Delta \bm p_{3,4} &-\Delta \bm p_{3,4} &\bm 0& \bm 0 &\bm 0\\
\Delta \bm p_{1,5} &\bm 0 &\bm 0 &\bm 0&-\Delta \bm p_{1,5} &\bm 0&\bm 0\\
\bm0 &\bm 0 &\bm 0 & \Delta \bm p_{4,5} &-\Delta \bm p_{4,5} &\bm 0&\bm 0\\
\bm 0& \Delta \bm p_{2,5} &\bm 0&\bm 0& -\Delta \bm p_{2,5}  &\bm 0&\bm 0\\
\bm 0& \bm 0&\Delta \bm p_{3,5} &\bm 0& -\Delta \bm p_{3,5}  &\bm 0&\bm 0\\
\Delta \bm p_{1,R_1}&\bm 0&\bm 0&\bm 0&\bm 0&-\Delta \bm p_{1,R_1}&\bm 0\\
\bm 0 &\Delta \bm p_{2,R_1}&\bm 0&\bm 0&\bm 0&-\Delta \bm p_{2,R_1}&\bm 0\\
\bm 0 &\bm 0&\Delta \bm p_{3,R_1} &\bm 0&\bm 0& -\Delta \bm p_{3,R_1}&\bm 0\\
\Delta \bm p_{1,R_2} &\bm 0&\bm 0&\bm 0&\bm 0&\bm0&-\Delta \bm p_{1,R_2} \\
\bm 0 &\bm 0&\bm 0&\Delta \bm p_{4,R_2} &\bm 0&\bm 0&-\Delta \bm p_{4,R_2}  \\
\bm 0 &\bm 0&\bm 0 &\bm 0& \Delta \bm p_{5,R_2} & \bm 0&-\Delta \bm p_{5,R_2}  
\end{array}
\right]}.
\label{3daxiom3a_2}
\end{multline}
We show in the following paragraph that the diagonal blocks: 
\begin{multline}
\scriptsize{
\left \{
\left[ \!\begin{array}{c}\Delta \bm p_{1,2} -\Delta \bm p_{1,2} \end{array}\! \right],\left[ \!\begin{array}{c} -\Delta \bm p_{1,3}  \\ -\Delta \bm p_{2,3} \end{array}\! \right],\left[ \!\begin{array}{c} -\Delta \bm p_{1,4}  \\ -\Delta \bm p_{2,4} \\ -\Delta \bm p_{3,4}  \end{array}\! \right],\left[ \!\begin{array}{c} -\Delta \bm p_{1,5}  \\ -\Delta \bm p_{4,5} \\ -\Delta \bm p_{2,5} \\ -\Delta \bm p_{3,5}  \end{array}\! \right],\left[ \!\begin{array}{c} -\Delta \bm p_{1,R_1} \\ -\Delta \bm p_{2,R_1} \\ -\Delta \bm p_{3,R_1}  \end{array}\! \right],\left[ \!\begin{array}{c} -\Delta \bm p_{1,R_2}  \\ - \Delta \bm p_{4,R_2} \\ -\Delta \bm p_{5,R_2}  \end{array}\! \right]
\right \}}
\label{3daxiom3a_3}
\end{multline}
have ranks $1$, $2$, $3$, $3$, $3$, and $3$, respectively. 

The first two and last two of these claims are trivial under the hypotheses that $\{\bm p_1,\bm p_2,\bm p_3\}$ and $\{\bm p_1,\bm p_4,\bm p_5\}$ are not rod-sharing sets. The third block would lose a dimension if some three-member subset of $\{\bm p_1, \bm p_2, \bm p_3, \bm p_4\}$ were collinear, but we now show they are not. First, $\{\bm p_1, \bm p_2, \bm p_3\}$ is assumed to be noncollinear. In addition, neither $\{\bm p_1, \bm p_2, \bm p_4\}$ nor $\{\bm p_1, \bm p_3, \bm p_4\}$ may be collinear under the generic conditions we have assumed, as each of these sets contains points from $R_1,$ $R_2$, and $R_3$. Because each of these sets contain two points on two separate rigid components (with one being an intersection between them), collinearity would imply that two distinct components contain at least one shared rod, which we exclude by construction. If $\{\bm p_2,\bm p_3, \bm p_4\}$ were collinear, then interchanging of $\bm p_4$ and $\bm p_5$ (which are undistinguishable in our hypothesis) would preserve the block rank of three, which we show now. Because $\{\bm p_2,\bm p_3,\bm p_4,\bm p_5\}$ is assumed to be noncollinear, collinearity of the set $\{\bm p_2,\bm p_3,\bm p_4\}$ guarantees noncollinearity of the set $\{\bm p_2,\bm p_3,\bm p_5\}$. As interchanging $\bm p_4$ and $\bm p_5$ does not affect the rank of the previously discussed blocks, we conclude that proper choice of $\bm p_4$ and $\bm p_5$ assures full rank of the third diagonal block. 

The fourth block has rank $\le 2$ only if either a four-member subset of $\{\bm p_1, \bm p_2, \bm p_3, \bm p_4, \bm p_5\}$ is collinear or two three-member sets of the involved constraints are both collinear. The first case is impossible, as $\{\bm p_2, \bm p_3, \bm p_4, \bm p_5\}$ is assumed to be noncollinear and any four-member set containing $\bm p_1$ includes points from all three components. Generically, if such a set were to be collinear, then the intersection of all three components would be nonempty---because intersections between (distinct) components are pairwise, this is impossible. This latter situation (that is, two collinear three-member subsets) is also impossible because $\Delta \bm p_{1,5}$ cannot generically be collinear with any of the other three constraints in the block, since it necessarily connects points on separate rods. This statement follows from the hypothesis that $\{\bm p_1, \bm p_4, \bm p_5\}$ is not collinear, and from the observation that neither $\{\bm p_1,\bm p_2, \bm p_5\}$ nor $\{\bm p_1,\bm p_3, \bm p_5\}$ can be collinear (for the same reason that neither $\{\bm p_1, \bm p_2, \bm p_4\}$ nor $\{\bm p_1, \bm p_3, \bm p_4\}$ can be collinear). 
Therefore, this fourth block has rank 3 and, because the rank of a block triangular matrix is bounded below by the sum of the ranks of its diagonal blocks, the composite rigidity matrix $\bm X_1\ast \bm X_2 \ast \bm X_3$ has rank $15$ and right nullspace has dimension six.

Finally, we turn to the case that any subset of the rigid bodies are planar. If either $R_1$ or $R_2$ is planar, then we choose minimal coordinate labelings appropriately (i.e., omitting $\bm p_{R_1}$ and/or $\bm p_{R_2}$ so as to keep only three points in the corresponding coordinate labeling). The proof proceeds equivalently, except that the rigidity matrix loses rows/columns that include $\bm p_{R_1}$ and/or $\bm p_{R_2}$. However, if $R_3$ is planar, then we must choose a non-minimal coordinate labeling for $R_3$, containing the four points $\{ \bm p_2, \bm p_3, \bm p_4, \bm p_5\}$. The resulting composite matrix has rank equivalence to the matrix in Eq.~\ref{3daxiom3a_1}, and the rank may be bounded by examining of the same diagonal blocks as above. Now, however, the third block in 
\ref{3daxiom3a_3} has rank $3$ in the generic case that $\bm p_1$ (which is not not contained in $R_3$) is non-coplanar with $R_3$. The fourth block also has rank 3: $\Delta \bm p_{4,5}$, $\Delta \bm p_{2,5}$, $\Delta \bm p_{3,5}$ are three parallel vectors spanning $\mathbb{R}^2$, but since $\Delta \bm p_{1,5}$ is (generically) non-coplanar with $R_3$ this block has rank 3.
\end{proof}

\begin{theorem}
(Rigidity of Motif 3D3B): If two nonaxisymmetric rigid bodies $R_1$ and $R_2$ ($n_r^1,n_r^2>1$) intersect at $\ge2$ instances (e.g., letting $\bm p_1,\bm p_2$ be points in two of the instances comprising $R_1\cup R_2$), and also each intersect another distinct rod $R_3$ ($n_r^3=1$) at exactly $1$ instance apiece, such that $\bm p_3\in R_1\cap R_3$ and $\bm p_4\in R_2\cap R_3$, then the composite body is rigid---so long as neither $\bm p_3$ nor $\bm p_4$ is collinear with both $\bm p_1$ and $\bm p_2$.
\end{theorem}
\begin{proof}
First, assume $R_1$ and $R_2$ are nonplanar. We choose as coordinate labelings $\{\bm p_1, \bm p_2, \bm p_3, \bm p_{R_1}\}$ for $R_1$, $\{\bm p_1, \bm p_2, \bm p_4, \bm p_{R_2}\}$ for $R_2$, and $\{\bm p_3, \bm p_4\}$ for $R_3$. We assume that $\bm p_{R_1}$ is not collinear with any pair of points in $\{\bm p_1, \bm p_2, \bm p_3\}$ and $\bm p_{R_2}$ to not be collinear with any pair of points in $\{\bm p_1, \bm p_2, \bm p_4\}$. Upon row rearrangement, this choice gives the rigidity matrix:
\begin{align}
\small{
\bm X_1 \ast \bm X_2 \ast \bm X_3=
\left[
\begin{array}{cccccc}
\Delta \bm p_{1,2} & -\Delta \bm p_{1,2} & \bm 0 &\bm 0 &\bm 0 &\bm 0\\
\Delta \bm p_{1,3} &\bm 0& - \Delta \bm p_{1,3} &\bm 0 &\bm 0 &\bm 0\\
\bm 0 & \Delta \bm p_{2,3} & - \Delta \bm p_{2,3} &\bm 0 &\bm 0 &\bm 0\\
\Delta \bm p_{1,4} &\bm 0  &\bm 0 & - \Delta \bm p_{1,4}&\bm 0 &\bm 0\\
\bm 0  &\Delta \bm p_{2,4} &\bm 0 & - \Delta \bm p_{2,4}&\bm 0 &\bm 0\\
\bm 0  &\bm 0 & \Delta \bm p_{3,4} &- \Delta \bm p_{3,4}&\bm 0 &\bm 0\\
\Delta \bm p_{1,R_1} & \bm 0 & \bm 0 &\bm 0 &-\Delta \bm p_{1,R_1}  &\bm 0\\
\bm 0 & \Delta \bm p_{2,R_1}& \bm 0 &\bm 0 &-\Delta \bm p_{2,R_1}  &\bm 0\\
\bm 0 & \bm 0& \Delta \bm p_{3,R_1} &\bm 0 &-\Delta \bm p_{3,R_1}  &\bm 0\\
\Delta \bm p_{1,R_2} & \bm 0 & \bm 0 &\bm 0  &\bm 0&-\Delta \bm p_{1,R_2} \\
\bm 0 & \Delta \bm p_{2,R_2}& \bm 0 &\bm 0   &\bm 0&-\Delta \bm p_{2,R_2}\\
\bm 0 & \bm 0&\bm 0 & \Delta \bm p_{4,R_2}  &\bm 0&-\Delta \bm p_{4,R_2} 
\end{array}
\right],}
\label{3daxiom3b_1}
\end{align}
which has diagonal blocks: 
\begin{equation}
\small{
\left \{
\left[ \begin{array}{cc} \Delta \bm p_{1,2} &-\Delta \bm p_{1,2} \end{array} \right],\left[ \begin{array}{c} -\Delta \bm p_{1,3}  \\ -\Delta \bm p_{2,3} \end{array} \right],\left[ \begin{array}{c} -\Delta \bm p_{1,4}  \\ -\Delta \bm p_{2,4} \\ -\Delta \bm p_{3,4}  \end{array} \right],\left[ \begin{array}{c} -\Delta \bm p_{1,R_1} \\ -\Delta \bm p_{2,R_1} \\ -\Delta \bm p_{3,R_1}  \end{array} \right],\left[ \begin{array}{c} -\Delta \bm p_{1,R_2}  \\ - \Delta \bm p_{2,R_2} \\ -\Delta \bm p_{4,R_2}  \end{array} \right]
\right \}.}
\end{equation}
Because we have chosen $\bm p_{R_1}$ and $\bm p_{R_2}$ to not be collinear with any pair of points in their respective coordinate labelings, and because of noncollinearity of the sets $\{\bm p_1, \bm p_2, \bm p_3\}$ and $\{\bm p_1, \bm p_2, \bm p_4\}$, the first, second, fourth, and fifth of these blocks trivially have full rank. To establish that the third block has full rank, we first claim that neither $\bm p_2$ nor $\bm p_1$ lie collinear with $\bm p_3$ and $\bm p_4$. Otherwise, one of $\bm p_2$ or $\bm p_1$ would lie within $R_3$. Because $R_1$ and $R_2$ intersect $R_3$ at only one instance apiece, this would mean that all $R_1,R_2,R_3$ intersect and so we could apply the special case noted in the second paragraph of Sec.~\ref{subsec:3dmotifs}. That is, we would choose $\bm p_1$ (or $\bm p_2$) explicitly so that it is non-collinear
with $\bm p_3$ and $\bm p_4$. The sets $\{\bm p_1, \bm p_2, \bm p_3\}$ and $\{\bm p_1, \bm p_2, \bm p_4\}$ are also non-collinear, implying that $\{\bm p_1, \bm p_2, \bm p_3, bm p_4\}$ is nonplanar and so the block consists of three vectors spanning $\mathbb{R}^3$ (i.e., it has full rank). Therefore $\bm X_1 \ast \bm X_2 \ast \bm X_3$ has rank $12$ and right nullspace dimension $6$. The case in which either $R_1$ or $R_2$ is planar proceeds similarly, choosing minimal coordinate labelings appropriately.
\end{proof}

\begin{theorem}
(Rigidity of Motif 3D4): If one nonaxisymmetric rigid body $R_1$ ($n_r^1>1$) intersects three rods $R_2,R_3,R_4$ ($n_r^2,n_r^3,n_r^4=1$) at the instances - each containing one of the points $\bm p_1, \bm p_2, \bm p_3$; $R_2$ intersects $R_3$ at one instance containing $\bm p_4$; and $R_3$ intersects $R_4$ at one instance containing $\bm p_5$, then the composite body is rigid unless $\bm p_1$, $\bm p_2$ and $\bm p_3$ lie along one rod.
\end{theorem}
\begin{proof}
We choose as minimal coordinate labelings $\{\bm p_1, \bm p_2 ,\bm p_3, \bm p_{R_1}\}$ for $R_1$ (which is assumed to be nonplanar)---where $\bm p_{R_1}$ is a point in $R_1$ that is not collinear with any pair of points in  $\{\bm p_1, \bm p_2 ,\bm p_3\}$; $\{ \bm p_1,\bm p_4\}$ for $R_2$; and $\{\bm p_3, \bm p_5\}$ for $R_4$. However, the rod $R_3$ contains three intersection points. Therefore, either pair can be chosen as an appropriate coordinate labeling (we choose $\{\bm p_4, \bm p_5\}$), but an augmented constraint must be added to enforce the condition that these three points must remain collinear (as in \emph{Motif 2D5} in \cite{rigidity_heroy}):
\begin{equation}
\frac{d\bm p_2}{dt}=s \frac{d\bm p_4}{dt}+(1-s)\frac{d\bm p_5}{dt},
\end{equation} 
where $s=\frac{||\Delta \bm p_{2,5}||_2}{||\Delta \bm p_{4,5}||_2}$. Letting $\bm {I_3}$ be the $3\times 3$ identity matrix, and $\bm {0_3}$ the $3\times 3$ all-zero matrix, these augmented constraints give the composite rigidity matrix:
\begin{align}
\small{
\bm X_1 \ast \bm X_2 \ast \bm X_3\ast \bm X_4=
\left[
\begin{array}{cccccc}
\Delta \bm p_{1,2} & -\Delta \bm p_{1,2} & \bm 0 &\bm 0 &\bm 0 &\bm 0\\
\Delta \bm p_{1,3} &\bm 0& -\Delta \bm p_{1,3} &\bm 0 &\bm 0 &\bm 0\\
\bm 0&\Delta \bm p_{2,3} & -\Delta \bm p_{2,3} &\bm 0 &\bm 0 &\bm 0\\
\Delta \bm p_{1,R_1} &\bm 0  &\bm 0 & - \Delta \bm p_{1,R_1}&\bm 0 &\bm 0\\
\bm 0  &\Delta \bm p_{2,R_1} &\bm 0 & - \Delta \bm p_{2,R_1}&\bm 0 &\bm 0\\
\bm 0 &\bm 0 &\Delta \bm p_{3,R_1}& - \Delta \bm p_{3,R_1}&\bm 0 &\bm 0\\
\Delta \bm p_{1,4} & \bm 0 & \bm 0 &\bm 0   &-\Delta \bm p_{1,4} &\bm 0\\
\bm 0 & \bm 0& \Delta \bm p_{3,5} &\bm 0   &\bm 0&-\Delta \bm p_{3,5}\\
\bm 0 & \bm 0& \bm 0   &\bm 0&\Delta \bm p_{4,5} &-\Delta \bm p_{4,5}\\
\bm {0_3} & \bm{I_3}&\bm{0_3}&\bm{0_3}&s\bm{I_3} &(1-s)\bm{I_3}
\end{array}
\right].
\label{3daxiom4_1}}
\end{align}
Elementary row operations give that this matrix is rank equivalent to: 
$$
\left[
\begin{array}{cccccc}
\Delta\bm p_{1,2} & -\Delta\bm p_{1,2} & \bm 0 &\bm 0 &\bm 0 &\bm 0\\
\Delta\bm p_{1,3} &\bm 0& -\Delta\bm p_{1,3} &\bm 0 &\bm 0 &\bm 0\\
\bm 0&\Delta\bm p_{2,3} & -\Delta\bm p_{2,3} &\bm 0 &\bm 0 &\bm 0\\
\Delta\bm p_{1,R_1} &\bm 0  &\bm 0 & - \Delta\bm p_{1,R_1}&\bm 0 &\bm 0\\
\bm 0  &\Delta\bm p_{2,R_1} &\bm 0 & - \Delta\bm p_{2,R_1}&\bm 0 &\bm 0\\
\bm 0 &\bm 0 &\Delta\bm p_{3,R_1}& - \Delta\bm p_{3,R_1}&\bm 0 &\bm 0\\
\Delta\bm p_{1,4} & \bm 0 & \bm 0 &\bm 0   &-\Delta\bm p_{1,4} &\bm 0\\
\bm 0 &  -\frac{1}{1-s} \Delta\bm p_{3,5}& \Delta \bm p_{3,5} &\bm 0   &\frac{s}{1-s}\Delta\bm p_{3,5}&\bm 0\\
\bm 0 & \frac{-1}{1-s} \Delta \bm p_{4,5}& \bm 0   &\bm 0&\left(1 +\frac{s}{1-s}\right) \Delta \bm p_{4,5}&\bm 0\\
\bm {0_3} & -\bm{I_3}&\bm{0_3}&\bm{0_3}&s\bm{I_3}&(1-s)\bm{I_3}
\end{array}
\right],
$$
which we claim has full rank diagonal blocks of rank $1,2,3,3,$ and $3$ (the first block being $1\times 6$, the second being $2\times 3$, and the rest being $3\times 3$). The first and last of these claims are trivial. Given that $\{\bm p_1, \bm p_2, \bm p_3, \bm p_{R_1}\}$  is chosen as a coordinate labeling with no three-member subset collinearities, the second and third blocks are full row rank. Finally, the fourth block is full rank because $\Delta \bm p_{1,4}$, $\Delta \bm p_{3,5}$, and $\Delta \bm p_{4,5}$ lie along three distinct (noncollinear) rods (and their coefficients are scalars).  This argument ensures that the matrix has rank at least $12$, and thus right nullspace dimension at most $6$. The proof in the case that $R_1$ is planar proceeds similarly.
\end{proof}
The final two motifs proven here involve individual rods only. The first (\emph{Motif 3D3C}) is obviously analogous to \emph{Motif 2D3} in \cite{rigidity_heroy}, although \emph{Motif 3D3B} is only applicable at the scale of single rods. Together, Motifs \emph{3D3C} and \emph{3D2B} show that a 3-clique community is also rigid in three dimensions. The last motif is analogous to \emph{Motif 2D5}, but an additional rod is needed to constrain the rigid motions of this structure in three dimensions. Before moving to the relevant theorems/proofs, we note that rigidity is a generic property of graphs that is generalizable to many different systems including central-force networks, body-hinge systems, and body bar networks \cite{asimow1978rigidity,whiteley1996some,Gluck}. While interested readers should refer in particular to \cite{asimow1978rigidity}, the argument in very nonrigorous terms may be stated as follows. If a rigidity matrix for a rigid body has rank $k(=3|V|-6$ in $D=3$), then the polynomial $P(\vec{ \bm p})$ in $D|V|$ variables given by the sum of determinants of all $k\times k$ submatrices of the rigidity matrix is nontrivial, and the set of ``regular points'', given by the positions of the vertices in any rigid configuration, form a dense open subset of $\mathbb{R}^{DN}$. Fubini's Theorem furthermore finally allows us to conclude that the set of singular points of $P$ has Lebesque measure zero. We could not apply this in the proofs above for various reasons amounting to either the individual components being of different scales (rods or nonaxisymmetric rigid bodies) or the need to attend to possibly singular cases in which contacts could be collinear. However, the two proofs below relate to individual rods.
\begin{theorem}
Motif 3D3C: A 3-clique of rods ($n_r^1,n_r^2,n_r^3=1$), wherein $R_1$ intersects $R_3$ at $\bm p_1$, $R_1$ intersects $R_2$ at one instance containing $\bm p_2$, and $R_2$ intersects $R_3$ at one instance containing $\bm p_3$, is rigid.
\end{theorem}
\begin{proof}
Choose as the coordinate labelings the respective intersection points. The resulting rigidity matrix is of size $3\times 9$ and trivially has full row rank.
\end{proof}

\begin{theorem}
(Rigidity of Motif 3D6): If six rods ($n_r^1=1,\cdots,n_r^6=1$) intersect in the strutted fashion of \emph{Motif 2D5} \cite{rigidity_heroy}---such that $\bm p_1\in (R_1\cap R_3)$, $\bm p_2\in (R_2\cap R_3)$, $\bm p_3\in (R_1\cap R_4)$, $\bm p_4\in (R_2\cap R_4)$, $\bm p_5\in (R_1\cap R_5)$, $\bm p_6\in (R_2\cap R_5)$, $\bm p_7\in (R_1\cap R_6)$, and $\bm p_8\in (R_2\cap R_6)$---then their composition is rigid.

\end{theorem}
We choose as minimal coordinate labelings $\{ \bm p_1, \bm p_7\}$ for $R_1$, $\{ \bm p_2, \bm p_8\}$ for $R_2$, $\{ \bm p_1, \bm p_2\}$ for $R_3$, $\{ \bm p_3, \bm p_4\}$ for $R_4$, $\{ \bm p_5, \bm p_6\}$ for $R_5$, and $\{ \bm p_7, \bm p_8\}$ for $R_6$. As in \emph{Motif 2D5} and \emph{Motif 3D4A}, we introduce augmented constraints to ensure that the positioning of $\bm p_3$ and $\bm p_5$ each stay fixed relative to $\bm p_1$ and $\bm p_7$ (and both $\bm p_4$ and $\bm p_6$ each stay fixed relative to $\bm p_2$ and $\bm p_8$) for all time.
\begin{multline} \bm X_1\ast... \ast \bm X_6= \\
\scriptsize{ \left[
\begin{array}{cccccccc}
\Delta \bm p_{1,2}&- \Delta \bm p_{1,2} &\bm 0 & \bm 0 & \bm 0& \bm 0& \bm 0& \bm 0 \\
\bm 0& \bm 0&\Delta \bm p_{3,4} & -\Delta \bm p_{3,4} & \bm 0&  \bm 0& \bm 0& \bm 0 \\
\Delta \bm p_{1,7} & \bm 0&\bm 0& \bm 0 & \bm 0& \bm 0& -\Delta \bm p_{1,7} & \bm 0 \\
\bm 0& \bm 0&\bm 0 & \bm 0 & \Delta \bm p_{5,6}& -\Delta \bm p_{5,6}& \bm 0& \bm 0\\
\bm 0& \Delta \bm p_{2,8}&\bm 0 & \bm 0& \bm 0 & \bm 0& \bm 0 &-\Delta \bm p_{2,8}\\
\bm 0& \bm 0 &\bm 0 & \bm 0 &  \bm 0& \bm 0 &\Delta \bm p_{7,8}& -\Delta \bm p_{7,8}\\
s_1\bm{I_3}&\bm{0_3} &-\bm{I_3} & \bm{0_3} & \bm{0_3}& \bm{0_3}&(1-s_1) \bm {I_3}& \bm{0_3} \\
\bm{0_3}&s_3\bm{I_3} &\bm{0_3} & -\bm{I_3} & \bm{0_3}& \bm{0_3}&  \bm{0_3}& (1-s_3)\bm {I_3}\\
s_2\bm{I_3}&\bm{0_3} &\bm{0_3} & \bm{0_3} & -\bm{I_3}& \bm{0_3}& (1-s_2)\bm {I_3}& \bm{0_3} \\
\bm{0_3}&s_4\bm{I_3} &\bm{0_3}  & \bm{0_3}& \bm{0_3}& -\bm{I_3}&  \bm{0_3} &(1-s_4)\bm {I_3}
\end{array}\right],}
\end{multline}
where $s_1=\frac{||\Delta \bm p_{1,3}||_2}{||\Delta \bm p_{1,7}||_2}$, $s_2=\frac{||\Delta \bm p_{1,5}||_2}{||\Delta \bm p_{1,7}||_2}$, $s_3=\frac{||\Delta \bm p_{2,4}||_2}{||\Delta \bm p_{2,8}||_2}$, $s_4=\frac{||\Delta \bm p_{2,6}||_2}{||\Delta \bm p_{2,8}||_2}$. We construct one (generic) realization for this system in $\mathbb{R}^3$ and show that it has full rank in silico, therefore concluding that any generic realization does as well.

%% file: algorithmicdetails.tex
\begin{figure}[htbp]
\begin{center}
\includegraphics[width=.8\linewidth]{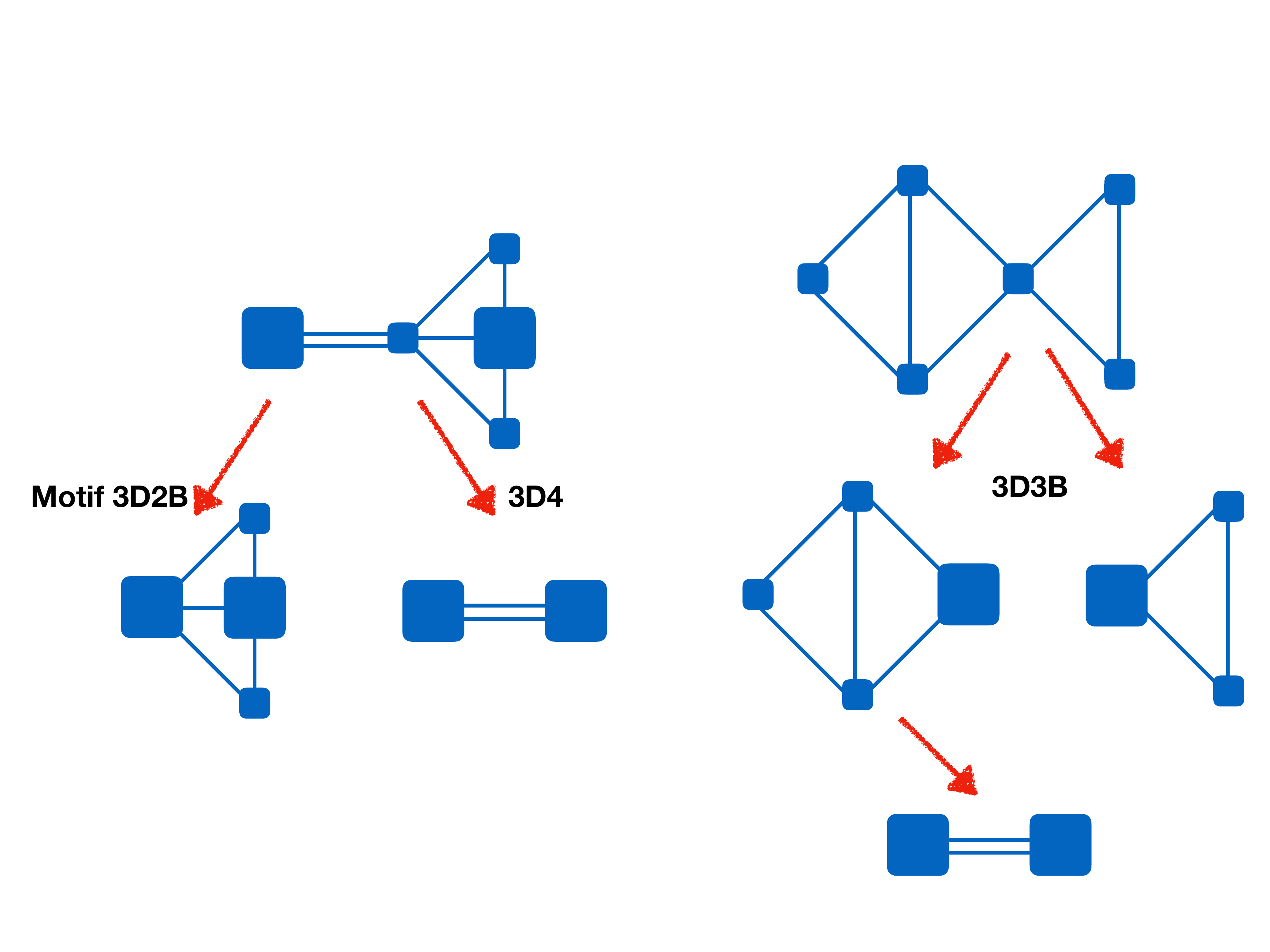}
\caption{Depending on which motif identification/compression is performed first on the top contact network representations, two different (irreducible) outcomes are possible are possible through \emph{3D-RGC} in each case. This comes about because our algorithm is based on the premise that rigid components ``come in one piece" (see Sec.~\ref{subsec:challenges}). However, we give evidence that the order of motif application does not largely affect our estimation of $\rho_{min,r}(\gamma)$ in Sec.~\ref{subsec:reshuffling}. }
\label{fig:order}
\end{center}
\end{figure}

In \cite{rigidity_heroy}, we showed that the ordering of motif compression was not consequential to the final results, as different orderings of compressing the three 2D rigid motifs in RGC always produced the same final state. However, as we show by counterexamples in Fig.~\ref{fig:order}, this is certainly not the case in 3 dimensions. Moreover, as further discussed in \cite{chubynsky2007algorithms}, any particle in a 2D rod-hinge system, or more generally a 2D central force network, may only be a part of one rigid component, wherein one particle may be part of many such components in a 3D rod-socket system. Given that our algorithm necessarily identifies each rod as being part of one component, we attempt to solve this problem by running \emph{3D-RGC} on the same networks many times in order to assess the global consistency of the algorithm, with the hypothesis that \emph{3D-RGC} will usually identify the largest rigid component in each system at least one of these times.  While we certainly cannot claim this hypothesis to be definitively true (there are a combinatorial number of ways to compress these large systems), we see in Sec.~\ref{subsec:reshuffling} that using one implementation is usually sufficient and using 10 implementations almost always achieves the same results as using 40. 

%% file: contact.tex
\begin{figure}
    \centering
    \includegraphics[width=\linewidth]{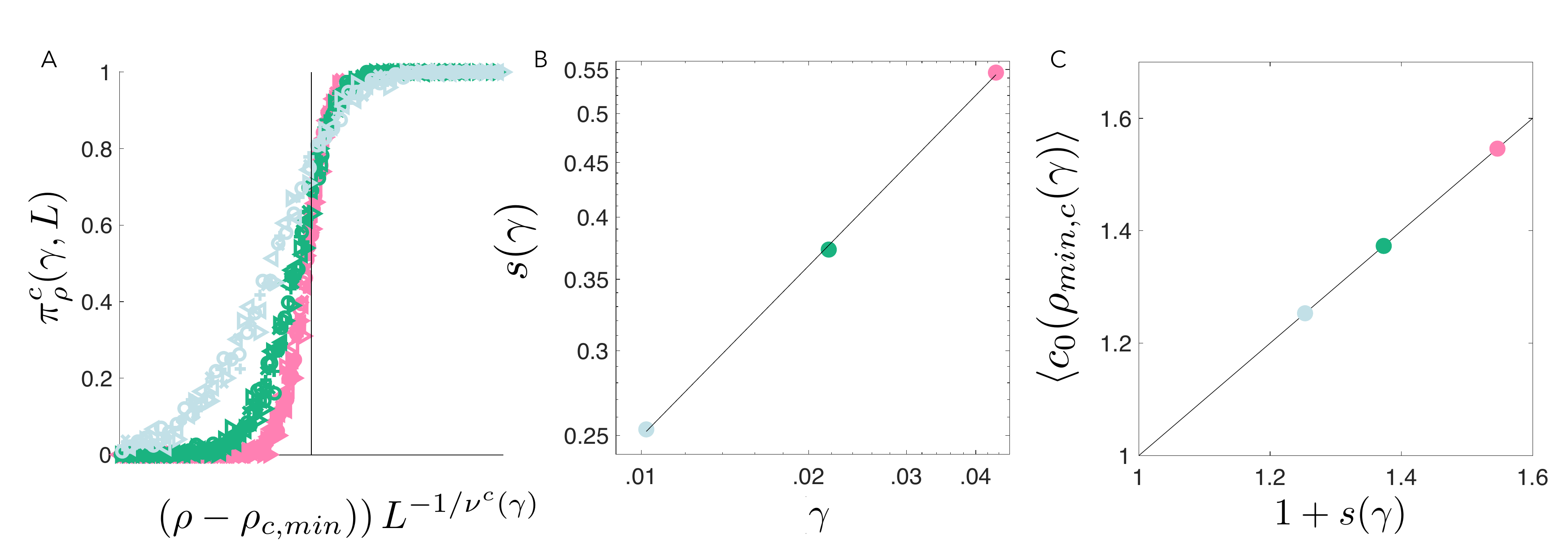}
    \caption{\bf{Recovery of contact percolation findings.} A. Scaling collapse for $\gamma=23,48,98$ (analogue of Eq.~\ref{eq:ansatz}). B. Fit of deviation parameter according to Eq.~\ref{eq:s}. C. Fit of mean critical degree against deviation parameter.}
    \label{fig:contact}
\end{figure}

In Fig.~\ref{fig:contact}, we display results for the detection of the contact percolation threshold, which we described in Secs.~\ref{subsec:exp},\ref{subsec:contact} and reported in Table~\ref{tab:2}. We also reproduce the scaling of $s$ with aspect ratio\footnote{Generally, these report a scaling with $r/\ell=1/(2\gamma)$, which we repeat for simplicity.} that has been observed in other studies, i.e.
\begin{equation}
s=c_1(1/2\gamma)^{c_2}.
\label{eq:s}
\end{equation}
We find via simple linear regression that $c_1=1.05$, $c_2=0.53$, and we note that other studies have sampled a wider range of $\gamma$ and gound $c_2\in[0.57,0.58]$ \cite{neda1999reconsideration,berhan2007modeling}).

%% file: 3drgcpaper.bbl
\begin{thebibliography}{10}

\bibitem{asimow1978rigidity}
{\sc L.~Asimow and B.~Roth}, {\em The rigidity of graphs}, Transactions of the
  American Mathematical Society, 245 (1978), pp.~279--289.

\bibitem{balberg1984percolation}
{\sc I.~Balberg, N.~Binenbaum, and N.~Wagner}, {\em Percolation thresholds in
  the three-dimensional sticks system}, Physical Review Letters, 52 (1984),
  p.~1465.

\bibitem{gorilla_glass}
{\sc P.~Ball}, {\em Material witness: Concrete mixing for gorillas}, Nature
  Materials, 14 (2015), p.~472.

\bibitem{berhan2007modeling}
{\sc L.~Berhan and A.~Sastry}, {\em Modeling percolation in high-aspect-ratio
  fiber systems. i. soft-core versus hard-core models}, Physical Review E, 75
  (2007), p.~041120.

\bibitem{broedersz2014modeling}
{\sc C.~P. Broedersz and F.~C. MacKintosh}, {\em Modeling semiflexible polymer
  networks}, Reviews of Modern Physics, 86 (2014), p.~995.

\bibitem{broedersz2011criticality}
{\sc C.~P. Broedersz, X.~Mao, T.~C. Lubensky, and F.~C. MacKintosh}, {\em
  Criticality and isostaticity in fibre networks}, Nature Physics, 7 (2011),
  p.~983.

\bibitem{bug1985continuum}
{\sc A.~Bug, S.~Safran, and I.~Webman}, {\em Continuum percolation of rods},
  Physical review letters, 54 (1985), p.~1412.

\bibitem{geometries}
{\sc J.~Cederberg}, {\em A course in modern geometries}, Springer, New York,
  2nd~ed., 2001.

\bibitem{celzard}
{\sc A.~Celzard, M.~Krzesi\~nska, J.~Mar\^eche, and F.~Puricelli}, {\em Scalar
  and vectorial percolation in compressed expanded graphite}, Physica A, 294
  (2001), pp.~283--94.

\bibitem{chaikin1995principles}
{\sc P.~M. Chaikin, T.~C. Lubensky, and T.~A. Witten}, {\em Principles of
  condensed matter physics}, vol.~10, Cambridge university press Cambridge,
  1995.

\bibitem{chatterjee2015random}
{\sc A.~P. Chatterjee and C.~Grimaldi}, {\em Random geometric graph description
  of connectedness percolation in rod systems}, Physical Review E, 92 (2015),
  p.~032121.

\bibitem{threedpebble}
{\sc M.~Chubynsky and M.~Thorpe}, {\em Algorithms for three-dimensional
  rigidity analysis and a first-order percolation transition}, Phys. Rev. E, 76
  (2007), p.~041135.

\bibitem{chubynsky2007algorithms}
{\sc M.~Chubynsky and M.~F. Thorpe}, {\em Algorithms for three-dimensional
  rigidity analysis and a first-order percolation transition}, Physical Review
  E, 76 (2007), p.~041135.

\bibitem{singer}
{\sc M.~Cucuringu, A.~Singer, and D.~Cowburn}, {\em Eigenvector
  synchronization, graph rigidity and the molecule problem}, Information and
  Inference, 1 (2012), pp.~21--67.

\bibitem{erdHos1960evolution}
{\sc P.~Erd{\H{o}}s and A.~R{\'e}nyi}, {\em On the evolution of random graphs},
  Publ. Math. Inst. Hung. Acad. Sci, 5 (1960), pp.~17--60.

\bibitem{favier1997mechanical}
{\sc V.~Favier, J.~Cavaille, G.~Canova, and S.~Shrivastava}, {\em Mechanical
  percolation in cellulose whisker nanocomposites}, Polymer Engineering \&
  Science, 37 (1997), pp.~1732--1739.

\bibitem{Gluck}
{\sc H.~Gluck}, {\em Almost all simply connected closed surfaces are rigid},
  Geometric Topology, Lecture Notes in Mathematics, 438 (1975), pp.~225--239.

\bibitem{graver}
{\sc J.~Graver}, {\em Rigidity matroids}, SIAM J. Discrete Math., 4 (1991),
  pp.~355--368, \url{https://doi.org/10.1137/0404032}.

\bibitem{glassy_metals}
{\sc P.~Gupta and D.~Miracle}, {\em A topological basis for bulk glass
  formation}, Acta Mater., 55 (2007), pp.~4507--15.

\bibitem{networkx}
{\sc A.~Hagberg, D.~Schult, and P.~Swart}, {\em Exploring network structure,
  dynamics, and function using networkx}, in Proceedings of the 7th Python in
  Science Conference (SciPy2008), G.~Varoquaux, T.~Vaught, and J.~Millman,
  eds., Scipy 2008, Pasadena, CA, 2008, pp.~11--15.

\bibitem{glasses_rigidity_diamond}
{\sc H.~He and M.~Thorpe}, {\em Elastic properties of glasses}, Phys. Rev.
  Lett., 54 (1985), pp.~2107--110.

\bibitem{hendrix}
{\sc B.~Hendrickson}, {\em Conditions for unique graph realizations}, SIAM J.
  Comput., 21 (1992), pp.~65--84,
  \url{https://doi.org/10.1016/j.compscitech.2011.04.010}.

\bibitem{henneberg1911graphische}
{\sc L.~Henneberg}, {\em Die graphische Statik der starren Systeme}, vol.~31,
  BG Teubner, 1911.

\bibitem{rigidity_heroy}
{\sc S.~Heroy, D.~Taylor, F.~B. Shi, M.~G. Forest, and P.~J. Mucha}, {\em Rigid
  graph compression: motif-based rigidity analysis for disordered fiber
  networks}, Multiscale Model. Simul., 16 (2018), pp.~1283--1304,
  \url{https://doi.org/10.1137/17M1157271}.

\bibitem{hespenheide2004structural}
{\sc B.~Hespenheide, D.~Jacobs, and M.~Thorpe}, {\em Structural rigidity in the
  capsid assembly of cowpea chlorotic mottle virus}, Journal of Physics:
  Condensed Matter, 16 (2004), p.~S5055.

\bibitem{hough2004viscoelasticity}
{\sc L.~Hough, M.~Islam, P.~Janmey, and A.~Yodh}, {\em Viscoelasticity of
  single wall carbon nanotube suspensions}, Physical review letters, 93 (2004),
  p.~168102.

\bibitem{lubensky}
{\sc M.~Huisman and T.~Lubensky}, {\em Internal stresses, normal modes, and
  nonaffinity in three-dimensional biopolymer networks}, Phys. Rev. Lett., 8
  (2011), p.~088301.

\bibitem{jackson2008pin}
{\sc B.~Jackson and T.~Jord{\'a}n}, {\em Pin-collinear body-and-pin frameworks
  and the molecular conjecture}, Discrete \& Computational Geometry, 40 (2008),
  pp.~258--278.

\bibitem{pebble}
{\sc D.~Jacobs and M.~Thorpe}, {\em Generic rigidity percolation: The pebble
  game}, Phys. Rev. Lett., 75 (1995), pp.~4051--4054.

\bibitem{jacobs1996generic}
{\sc D.~Jacobs and M.~Thorpe}, {\em Generic rigidity percolation in two
  dimensions}, Physical Review E, 53 (1996), p.~3682.

\bibitem{jacobs1998generic}
{\sc D.~J. Jacobs}, {\em Generic rigidity in three-dimensional bond-bending
  networks}, Journal of Physics A: Mathematical and General, 31 (1998),
  p.~6653.

\bibitem{jacobs2001protein}
{\sc D.~J. Jacobs, A.~J. Rader, L.~A. Kuhn, and M.~F. Thorpe}, {\em Protein
  flexibility predictions using graph theory}, Proteins: Structure, Function,
  and Bioinformatics, 44 (2001), pp.~150--165.

\bibitem{oxide_1}
{\sc R.~Kerner and J.~Phillips}, {\em Quantitative principles of silicate glass
  chemistry}, Solid State Commun., 117 (2000), pp.~47--51.

\bibitem{laman}
{\sc G.~Laman}, {\em On graphs and rigidity of plane skeletal structures}, J.
  Eng. Math., 4 (1970), pp.~331--340.

\bibitem{rigidity_fibers2d}
{\sc M.~Latva-Kokko and J.~Timonen}, {\em Rigidity of random networks of stiff
  fibers in the low-density limit}, Phys. Rev. E, 64 (2001), p.~066117.

\bibitem{gorilla_glass2}
{\sc J.~Mauro, A.~Ellison, and L.~Pye}, {\em Glass: the nanotechnology
  connection}, Int. J. Appl. Glass Sci., 4 (2013), pp.~64--75.

\bibitem{maxwell}
{\sc J.~Maxwell}, {\em On the calculation of the equilibrium and stiffness of
  frames}, Philosophical Magazine, 27 (1864), pp.~294--299.

\bibitem{graphlet}
{\sc B.~McKay}, {\em Graphs}.
\newblock
  \href{http://users.cecs.anu.edu.au/~bdm/data/graphs.html}{http://users.cecs.anu.edu.au/~bdm/data/graphs.html}.

\bibitem{neda1999reconsideration}
{\sc Z.~Neda, R.~Florian, and Y.~Brechet}, {\em Reconsideration of continuum
  percolation of isotropically oriented sticks in three dimensions}, Physical
  Review E, 59 (1999), p.~3717.

\bibitem{newman2009random}
{\sc M.~E. Newman}, {\em Random graphs with clustering}, Physical review
  letters, 103 (2009), p.~058701.

\bibitem{cpm}
{\sc G.~Palla, I.~Der{\'e}nyi, I.~Farkas, and T.~Vicsek}, {\em Uncovering the
  overlapping community structure of complex networks in nature and society},
  Nature Letters, 435 (2005), pp.~814--818.

\bibitem{pardalos2018open}
{\sc P.~M. Pardalos and A.~Migdalas}, {\em Open Problems in Optimization and
  Data Analysis}, vol.~141, Springer, 2018.

\bibitem{pellegrino1993structural}
{\sc S.~Pellegrino}, {\em Structural computations with the singular value
  decomposition of the equilibrium matrix}, International Journal of Solids and
  Structures, 30 (1993), pp.~3025--3035.

\bibitem{penu}
{\sc C.~Penu, G.~Hu, A.~Fernandez, P.~Marchal, and L.~Choplin}, {\em
  Rheological and electrical percolation thresholds of carbon nanotube/polymer
  nanocomposites}, Polym. Eng. Sci., 52 (2012), pp.~283--94.

\bibitem{philipse1996random}
{\sc A.~P. Philipse}, {\em The random contact equation and its implications for
  (colloidal) rods in packings, suspensions, and anisotropic powders},
  Langmuir, 12 (1996), pp.~1127--1133.

\bibitem{chalcogenide}
{\sc J.~Phillips}, {\em Topology of covalent non-crystalline solid i:
  short-range order in chalcogenide alloys}, J. Non-Cryst. Solids, 34 (1979),
  pp.~153--81.

\bibitem{proteins_1}
{\sc J.~Phillips}, {\em Constraint theory and hierarchical protein dynamics},
  J. Phys.: Condens. Matter, 16 (2004), pp.~S5065--72.

\bibitem{proteins_2}
{\sc J.~Phillips}, {\em Scaling and self-organized criticality in proteins:
  lysozyme c}, Phys. Rev. E, 80 (2009), p.~051916.

\bibitem{oxide_2}
{\sc J.~Phillips and R.~Kerner}, {\em Structure and function of window glass
  and pyrex}, J. Chem. Phys., 128 (2008), p.~174506.

\bibitem{constraint_theory}
{\sc J.~Phillips and M.~Thorpe}, {\em Constraint theory, vector percolation and
  glass formation}, Solid State Commun., 53 (1985), pp.~699--702.

\bibitem{schilling2015percolation}
{\sc T.~Schilling, M.~A. Miller, and P.~Van~der Schoot}, {\em Percolation in
  suspensions of hard nanoparticles: From spheres to needles}, EPL (Europhysics
  Letters), 111 (2015), p.~56004.

\bibitem{chalcogenides_4}
{\sc U.~Senapati and A.~Varshneya}, {\em Configurational arrangements in
  chalcogenide glasses: a new perspective on phillips? constraint theory}, J.
  Non-Cryst. Solids, 185 (1995), pp.~289--96.

\bibitem{scaling}
{\sc F.~Shi, S.~Wang, M.~Forest, and P.~Mucha}, {\em Percolation-induced
  exponential scaling in the large current tails of random resistor networks},
  Multiscale Model. Simul., 11 (2013), pp.~1298--1310,
  \url{https://doi.org/10.1137/130914929}.

\bibitem{shi}
{\sc F.~Shi, S.~Wang, M.~Forest, and P.~Mucha}, {\em Network-based assessments
  of percolation-induced current distributions in sheared rod macromolecular
  dispersions}, Multiscale Model. Simul., 12 (2014), pp.~249--264,
  \url{https://doi.org/10.1137/130926390}.

\bibitem{silva}
{\sc J.~Silva, R.~Simoes, S.~Lancos-Mendez, and R.~Vaia}, {\em Applying complex
  network theory to the understanding of high-aspect-ratio carbon-filled
  composites}, Europhys. Lett., 93 (2011), p.~37005.

\bibitem{sitharam2004tractable}
{\sc M.~Sitharam and Y.~Zhou}, {\em A tractable, approximate, combinatorial 3d
  rigidity characterization}, Fifth Automated Deduction in Geometry (ADG),
  (2004).

\bibitem{chalcogenides_3}
{\sc A.~Sreenum, D.~Swiler, and A.~Varshneya}, {\em Gibbs-dimarzio equation to
  describe the glass transition temperature trends in multicomponent
  chalcogenide glasses}, J. Non-Cryst. Solids, 127 (1991), pp.~287--97.

\bibitem{stauffer2018introduction}
{\sc D.~Stauffer and A.~Aharony}, {\em Introduction to percolation theory}, CRC
  press, 2018.

\bibitem{chalcogenides_2}
{\sc D.~Swiler, A.~Varshneya, and R.~Callahan}, {\em Microhardness, surface
  toughness and average coordination number in chalcogenide glasses}, J.
  Non-Cryst. Solids, 125 (1990), pp.~250--57.

\bibitem{tay1985generating}
{\sc T.-S. Tay and W.~Whiteley}, {\em Generating isostatic frameworks},
  Structural Topology 1985 N{\'u}m 11,  (1985).

\bibitem{glasses_rigidity_percolation}
{\sc M.~Thorpe}, {\em Rigidity percolation in glassy structures}, J. Non-Cryst.
  Solids, 76 (1985), pp.~109--16.

\bibitem{thorpe1999rigidity}
{\sc M.~F. Thorpe and P.~M. Duxbury}, {\em Rigidity theory and applications},
  Springer Science \& Business Media, 1999.

\bibitem{whiteley1996some}
{\sc W.~Whiteley}, {\em Some matroids from discrete applied geometry},
  Contemporary Mathematics, 197 (1996), pp.~171--312.

\bibitem{whiteley1999rigidity}
{\sc W.~Whiteley}, {\em Rigidity of molecular structures: geometric and generic
  analysis, rigidity theory and applications (edited by m. thorpe and p.
  duxbury)}, 1999.

\bibitem{wilhelm2003elasticity}
{\sc J.~Wilhelm and E.~Frey}, {\em Elasticity of stiff polymer networks},
  Physical review letters, 91 (2003), p.~108103.

\end{thebibliography}
